\documentclass[10pt,journal,compsoc]{IEEEtran}
\ifCLASSOPTIONcompsoc
  \usepackage[nocompress]{cite}
\else
  \usepackage{cite}
\fi

\ifCLASSINFOpdf
\else
\fi

\usepackage{amssymb}
\usepackage{amsmath}
\usepackage{graphicx}
\usepackage{multicol}

\DeclareMathOperator*{\argmax}{arg\,max}
\DeclareMathOperator*{\argmin}{arg\,min}

\begin{document}
%
\title{SCAUL: Power Side-Channel Analysis with Unsupervised Learning}

\author{Keyvan~Ramezanpour,
        Paul~Ampadu,
        and~William~Diehl
\IEEEcompsocitemizethanks{\IEEEcompsocthanksitem Authors are with the Department
of Electrical and Computer Engineering, Virginia Tech, Blacksburg,
VA, 24061.\protect\\
E-mail: \{rkeyvan8, ampadu, wdiehl\}@vt.edu.}
}


\IEEEtitleabstractindextext{%
\begin{abstract}
Existing power analysis techniques rely on strong adversary models with prior knowledge of the leakage or training data. We introduce side-channel analysis with unsupervised learning (SCAUL) that can recover the secret key without requiring prior knowledge or profiling (training). We employ an LSTM auto-encoder to extract features from power traces with high mutual information with the data-dependent samples of the measurements. We demonstrate that by replacing the raw measurements with the auto-encoder features in a classical DPA attack, the efficiency, in terms of required number of measurements for key recovery, improves by 10X. Further, we employ these features to identify a leakage model with sensitivity analysis and multi-layer perceptron (MLP) networks. SCAUL uses the auto-encoder features and the leakage model, obtained in an unsupervised approach, to find the correct key. On a lightweight implementation of AES on Artix-7 FPGA, we show that SCAUL is able to recover the correct key with 3700 power measurements with random plaintexts, while a DPA attack requires at least 17400 measurements. Using misaligned traces, with an uncertainty equal to 20\% of the hardware clock cycle, SCAUL is able to recover the secret key with 12300 measurements while the DPA attack fails to detect the key.
\end{abstract}

}

\maketitle

\IEEEdisplaynontitleabstractindextext

\IEEEpeerreviewmaketitle

\IEEEraisesectionheading{\section{Introduction}\label{sec:introduction}}

\IEEEPARstart{S}{ide}-Channel Analysis (SCA) using power consumption or electromagnetic (EM) emanations from electronic devices is a powerful tool for inferring information about hardware/software characteristics and processed data in a computing platform. Side-channel analysis refers to a technique in which behavior of a computing platform, including power consumption, EM radiation, timing and memory access, are observed to retrieve secret information. An SCA attack that analyzes the power traces or EM signals is usually referred to as power/EM analysis. 

Power analysis (PA) has especially been employed to compromise the security of different crypto-systems running on a computing platform. Examples include secret key recovery from elliptic-curve cryptography (ECC) running on iOS and Android devices \cite{genkin2016ecdsa} and McEliece cryptosystem implemented on FPGA \cite{chen2015horizontal}, attacks on Xilinx bitstream encryption \cite{moradi2016improved}, recovering the secret key of postquantum key exchange protocols \cite{aysu2018horizontal, kramer2018differential}, key recovery of Advanced Encryption Standard (AES) \cite{mangard2005successfully}, symmetric encryption systems \cite{heuser2016side, moos2019static} and breaking the security of smart cards \cite{mahanta2015power}. 

Existing power analysis techniques can be categorized into two groups of model-based and profiling attacks. In model-based attacks, prior knowledge of the leakage model is assumed that defines a relationship between the power consumption of a device and the processed data. In differential power analysis (DPA) \cite{kocher1999differential}, the measurements are clustered into two or more classes of similar traces, according to the leakage model. Statistics of the traces, e.g. the mean of power samples in first-order DPA, represent the traces in each cluster. The inter-cluster difference of the statistics is used as a measure to identify the correct data. In correlation power analysis (CPA) \cite{brier2004correlation}, the correlation coefficient between the power samples and the leakage model is used as the statistic to identify the correct data.

A commonly used leakage model is Hamming Weight (HW), according to which the power consumption of a logic block is correlated with the HW of the processed data \cite{alioto2009leakage}. Hamming Distance (HD) is also a popular model for power consumption corresponding to memory transitions, e.g. registers of microprocessors, in which the power is correlated with the HD between the initial and final values of memory elements \cite{coron2012conversion}.
Additionally, particular features of measured power traces might also be correlated with single bits of the data, e.g. the most significant bit (MSB) as used in \cite{timon2019non}. Switching glitches in hardware implementation of logic functions and toggling activity of internal nodes of the circuit are also shown to depend on data \cite{mangard2005successfully, sadhukhan2019count}.

Model-based power analysis relies on a significant amount of prior knowledge of the details of the hardware including the hardware architecture, CMOS technology, the specific implementation of cipher operations, power delivery circuitry, and even the layout of interconnects in integrated circuits \cite{das2019stellar}. Profiling techniques use actual power measurements of a device corresponding to known processed data to develop more accurate leakage models. In \cite{schindler2005stochastic}, a stochastic model is employed in which a polynomial function of data with random coefficients represents the mean of the leakage signal assumed to follow a Gaussian distribution. The coefficients are estimated using linear or ridge regression \cite{schindler2005stochastic, wang2017ridge}.

Machine learning has been employed to develop leakage models in a profiling-based approach. Support vector regression (SVR) is used in \cite{jap2015support} to develop a mapping from the bitwise representation of data to the power samples, assumed to follow a multivariate Gaussian distribution. In \cite{yang2011back}, a multi-layer perceptron (MLP) neural network has been employed, as a generic nonlinear leakage model, in which the inputs are the bits of data and the outputs are the mean of power samples corresponding to the data values. A partition-based approach is introduced in \cite{whitnall2015robust}, in which a set of power traces corresponding to known data are clustered using an unsupervised clustering algorithm such as k-means. The clusters constitute the leakage model; the sets of data at each cluster have similar power consumption.

Whether a leakage model is obtained a priori or through profiling, power analysis also requires suitable statistics of the power traces as distinguishers. Classical techniques such as DPA and CPA use predefined statistics such as first or higher order moments and the correlation coefficient of individual power samples with a leakage model. Mutual information and Kolmogorov–Smirnov statistics are also used as alternative distinguishers \cite{gierlichs2008mutual, whitnall2011exploration}. More recent techniques use profiling to extract suitable statistics. In template attacks \cite{chari2002template}, multivariate Gaussian distribution is assumed for the power traces. The mean and covariance of the distribution depends on the data and are estimated in a profiling step. More advanced techniques exploit supervised learning algorithms, such as support vector machine (SVM), decision tree (DT) and random forest (RF) \cite{lerman2011side}, and deep learning \cite{maghrebi2016breaking, timon2019non}, to extract the most relevant features of power traces, and develop proper distinguishers.

While profiling and supervised learning techniques are the most powerful SCA attacks on cryptographic implementations, their success rate rapidly degrades if the training set, captured during profiling on a reference device, slightly deviates from the measurements on the target device under attack. It is shown in \cite{wang2019diversity} that the accuracy of an MLP neural network in attacking AES running on an ATxmega128D4 microcontroller drops from 88.5\% to less than 13.7\% if the MLP is trained with power measurements on one board, and used to attack the same microcontroller, but on a different board. Having access to the identical hardware as the target is a major limitation of profiling-based techniques.

In this work, we introduce an unsupervised learning technique for side-channel analysis, called SCAUL, which does not require any training set for profiling or a prior knowledge of the leakage model. We employ a Long Short-Term Memory (LSTM) auto-encoder to extract features from power traces. The power features provide a similarity model between the power traces. An MLP neural network is used to map the power features to the processed data for a key candidate. Using \textit{sensitivity} analysis, a leakage model is identified, which is the characteristic of the similarity model extracted by the auto-encoder. The power features are clustered based on the identified leakage model, and the correct key exhibits the maximum inter-cluster difference. We demonstrate the success of SCAUL on a lightweight implementation of AES on FPGA, even with non-aligned power traces.

\begin{figure*}[htbp]
	\centering
	\begin{multicols}{2}
		\includegraphics[width=1\textwidth]{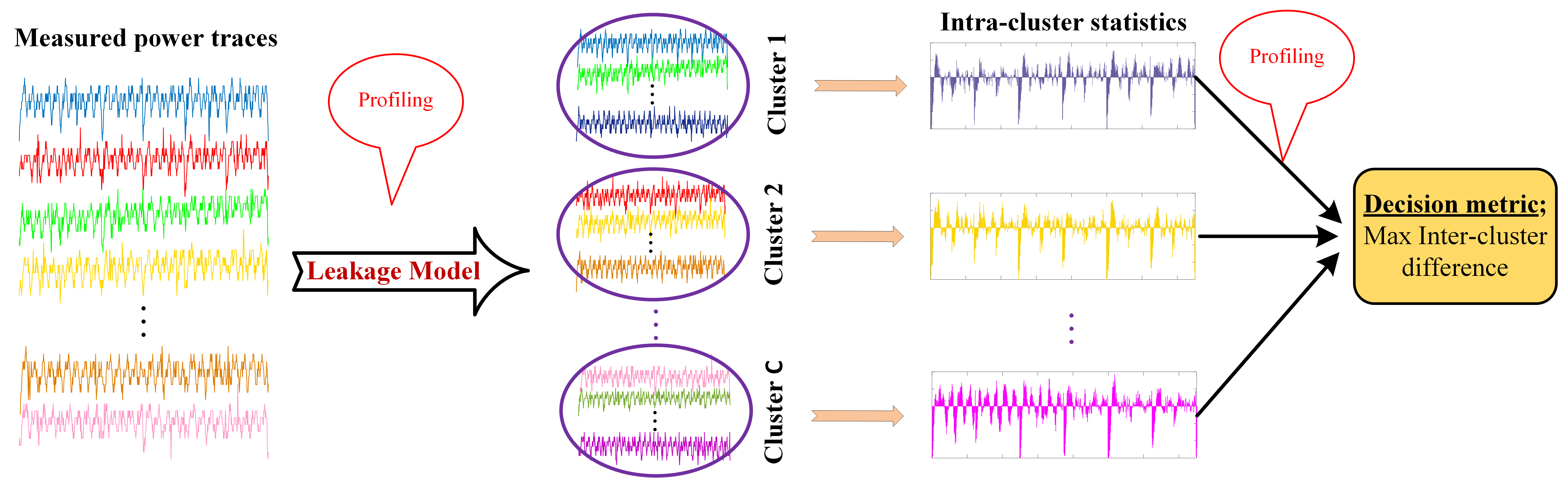}
	\end{multicols}
	\vspace{-0.6cm}
	\caption{Conceptual description of model-based power analysis techniques using a leakage model.}
	\label{fig:modelbased}
\end{figure*}

The contributions of this work include: 1) We introduce an LSTM auto-encoder that extracts data-dependent features from measurements in an unsupervised approach. It allows for a horizontal processing of power traces which improves the efficiency of an attack significantly. 2) We develop sensitivity analysis with MLP neural networks to detect a leakage model in an unsupervised approach. 3) We introduce the SCAUL technique for power analysis which recovers the secret key without requiring prior knowledge on a leakage model or profiling, even when data-dependent features are distributed over random time samples.

The rest of the paper is organized as follows. In Section \ref{sec:back}, an overview of different classes of existing power analysis techniques is provided and the mathematical background on the proposed SCAUL attack is developed. Section \ref{sec:method} describes the overall methodology of SCAUL, including the LSTM neural network for feature extraction and sensitivity analysis for leakage detection. A case study of the SCAUL attack on an FPGA implementation of AES is discussed in Section \ref{sec:AES}. Experimental results are presented in Section \ref{sec:results} and the paper concludes in Section \ref{sec:conclusion}.

\section{Background} \label{sec:back}
\subsection{Attack Model}
A general model for power analysis involves a key-dependent cipher operation $F_k(): \mathbb{F}_2^m \to \mathbb{F}_2^n$, a known input data to the operation as $Z \in \mathbb{F}_2^m$ and an unknown variable at the output of the operation as $X \in \mathbb{F}_2^n$ called the intermediate or sensitive variable. In most block ciphers the operation under attack is the nonlinear S-box function $S()$ while $m$ and $n$ are the number of bits at the input and output of the operation. The input $Z$ and the intermediate variable $X$ are $m$-bit subsets of the input plaintext and the cipher state, respectively. Further, the secret parameter $k$ is an $m$-bit subset of the entire secret key. Under this model, the S-box operation can be represented as $X=F_k(Z)=S(Z\oplus k)$, in which $\oplus$ is the bitwise XOR operation.

The fundamental property of a cipher operation exploited in power analysis to detect the secret key is \textit{independence} of the output bits of the operation from the input. Formally, using the binary representation of the intermediate variable $X$ as $\Bar{\mathrm{\mathbf{x}}}=(x_i)_{i=0,1,\cdots,m}$, we have $H(\Bar{\mathrm{\mathbf{x}}}_r|Z) = H(\Bar{\mathrm{\mathbf{x}}}_r)$, in which $H()$ is the Shannon Entropy and $\Bar{\mathrm{\mathbf{x}}}_r$ is any combination of $r\in [1,m]$ bits of $\Bar{\mathrm{\mathbf{x}}}$. However, $H(\Bar{\mathrm{\mathbf{x}}}_r|Z, k) = 0$; i.e., having the secret key, the cipher operation is a deterministic relation while without knowledge on $k$, the operation is a random transformation. In power analysis, the power consumption of the hardware implementation of the cipher operation is given as a vector of $N$ samples denoted by $\mathrm{\mathbf{T}}\in \mathbb{R}^N$. It is assumed that $I(\mathrm{\mathbf{T}}; \Bar{\mathrm{\mathbf{x}}}) > 0$, in which $I(a;b)$ is the mutual information between random variables $a$ and $b$.

Using the above properties, the primary idea of a power analysis technique is as follows. Having a set of input values to the cipher operation and the corresponding power traces during execution of the operation, an attacker calculates the values of the intermediate variable for all possible values of the secret key $k$. Let $\Bar{\mathrm{\mathbf{x}}}_{k^*}=F_{k^*}(Z)$ denote the output of the cipher operation with the input $Z$ and a key candidate $k^*$. If $k^*$ is the correct key, $I(\mathrm{\mathbf{T}}; \Bar{\mathrm{\mathbf{x}}}_{k^*}) > 0$ since $I(\mathrm{\mathbf{T}}; \Bar{\mathrm{\mathbf{x}}}) > 0$, otherwise, $I(\mathrm{\mathbf{T}}; \Bar{\mathrm{\mathbf{x}}}_{k^*}) = 0$. Hence, the mutual information between the power traces and the intermediate variable calculated for a key candidate can be considered as a metric to rank key candidates. The highest rank belongs to the correct key.

To implement a power attack in practice, the mutual information between the intermediate variable and power traces is usually expressed in the form of a leakage model. Let the function $L(): \mathbb{F}_2^m \to \mathbb{R}^N$ denote the leakage model which maps the $m$-bit intermediate variable $X$ to the vector of power trace $\mathrm{\mathbf{T}}$ with $N$ samples. A generic nonlinear leakage model can be expressed as the following algebraic normal form \cite{wang2017ridge}
\begin{equation} \label{eq:leakmodel}
    \Tilde{\mathrm{\mathbf{T}}} = L(X) = \alpha_0 + \sum_{U\in \mathbb{F}_2^m\backslash \{0\}} \alpha_U X^U + \epsilon
\end{equation}
in which $\Tilde{\mathrm{\mathbf{T}}}$ is the data-dependent component of the leakage power, $\epsilon$ is a noise and $X^U=\prod_{i=1}^m x_i^{u_i}$ is a monomial of degree $d=HW(U)$ representing the product of bits of $X$ at the positions where the corresponding bits of $U$ are $1$. 
\subsection{Model-Based Attacks}
In classical model-based power analysis, including DPA and CPA, it is assumed that there is at least one sample of the power trace $\mathrm{\mathbf{T}}$ which is correlated with the intermediate variable $X$ according to (\ref{eq:leakmodel}), with fixed coefficients $\alpha_U$ for all samples. The power sample which exhibits the highest correlation with the leakage model is chosen as the point of interest (POI) for ranking the key candidates. In profiling-based attacks, different coefficients might be estimated for every sample of the power trace. 

The overall procedure of a model-based attack is shown in Fig. \ref{fig:modelbased}. A set of $S$ power traces each with $N$ samples, denoted by $\mathrm{\mathbf{T}}_j=(t_{j,1},t_{j,2},\cdots,t_{j,N}), j=1,2,\cdots,S$, with the corresponding inputs $Z_j$ to a cipher operation $F_k()$ is available. For every key candidate $k^*$, the corresponding intermediate variable $X$ for a power trace $j$ is calculated as $X_{j, k^*}=F_{k^*}(Z_j)$. The power traces are grouped into $C$ clusters, in general, according to the calculated intermediate values and the leakage model. For example, in an HW model, it is assumed that the power traces corresponding to the intermediate variables with the same HW are similar. Hence, the clusters are $\mathbb{H}_c=\{X_{j,k^*}|HW(X_{j,k^*})=c\}$ with $c\in \{0,1,2,\cdots,m\}$ for an $m$-bit $X$, and $HW(X)=\sum_{i=0}^{m-1}x_i$ is the Hamming weight of $X$.

In classical techniques, statistics of the power traces in a cluster are calculated for every sample of the trace. In first order DPA, the mean of samples of power traces, i.e. $\Bar{\mathrm{\mathbf{T}}}_c=(\Bar{t}_{c,1},\Bar{t}_{c,2},\cdots,\Bar{t}_{c,N})$, is used as the cluster statistic, in which $\Bar{t}_{c,n}=\mathrm{E}_{X_{j,k^*}\in \mathbb{H}_c}[t_{j,n}], n=1,2,\cdots,N$. In a difference of means (DoM) test, the difference between the means of power samples between any two combination of clusters is used as the statistic to rank the key candidate $k^*$ \cite{standaert2008partition}; the correct key exhibits the maximum difference. Alternatively, in mutual information analysis (MIA), the mutual information between the measurements and the model is used as the decision metric to rank the key candidates \cite{batina2011mutual}. Higher order cluster statistics can also be used to attack low-order masked implementations as in \cite{prouff2009statistical} and \cite{schneider2015leakage}. The profiling-based leakage model can also be used in classical DPA attacks for the purpose of clustering as exploited in \cite{wang2017ridge, whitnall2015robust}.

Rather than clustering, in correlation power analysis (CPA), a mathematical model, as in (\ref{eq:leakmodel}), is used to characterize the data-dependent leakage \cite{biryukov2016correlation, repka2015correlation}. Thus, CPA can be considered as a generalization of DPA and MIA in which the number of clusters is equal to the space size of the intermediate variable $X$. The Pearson's correlation coefficient between the measured power and the estimated leakage according to the model is used to rank the key candidates. The coefficients of the leakage model can be assumed a priori, as in an HW or HD model, or obtained through profiling in a stochastic model. 

Profiling techniques are also used in identifying proper cluster statistics and decision metrics in model-based attacks. A popular profiling power analysis is template attack (TA) in which the probability density function (pdf) of a cluster is estimated in a \textit{profiling phase}, given a set of power traces corresponding to \textit{known} intermediate variables \cite{choudary2013efficient}. These traces should be collected from the same hardware, with known secret key, as the device under attack. The pdf of clusters are called \textit{templates} whose parameters depend on data. During the \textit{attack phase}, a measured power trace, from the device with unknown secret key, is matched with the templates using decision statistics such as maximum likelihood (ML) or Bayesian statistics, i.e. maximum a posteriori (MAP) estimation, to estimate the value of the intermediate variable. Profiling can be used for both leakage modeling and distribution estimation as in \cite{jap2015support}.

Classical model-based power analysis techniques assume aligned measurements in which the data-dependent features of power traces always appear at the same time sample which is a major limitation for two reasons: 1) timing of the measurements might not be precise relative to the timing of the device under attack; and 2) simple countermeasures such as addition of random clock jitters can result in the attack failure. Deep learning has been employed to address the issue of misaligned power traces. In \cite{timon2019non}, a convolutional neural network (CNN) is used to classify power traces according to a leakage model, including HW and MSB, for every key candidate. By adding a small perturbation to the weights of a trained neural network for a key candidate, the sensitivity of cluster probabilities to the perturbation is calculated. The correct key exhibits the highest sensitivity.

\begin{figure}[t!]
	\centering
		\includegraphics[width=0.5\textwidth]{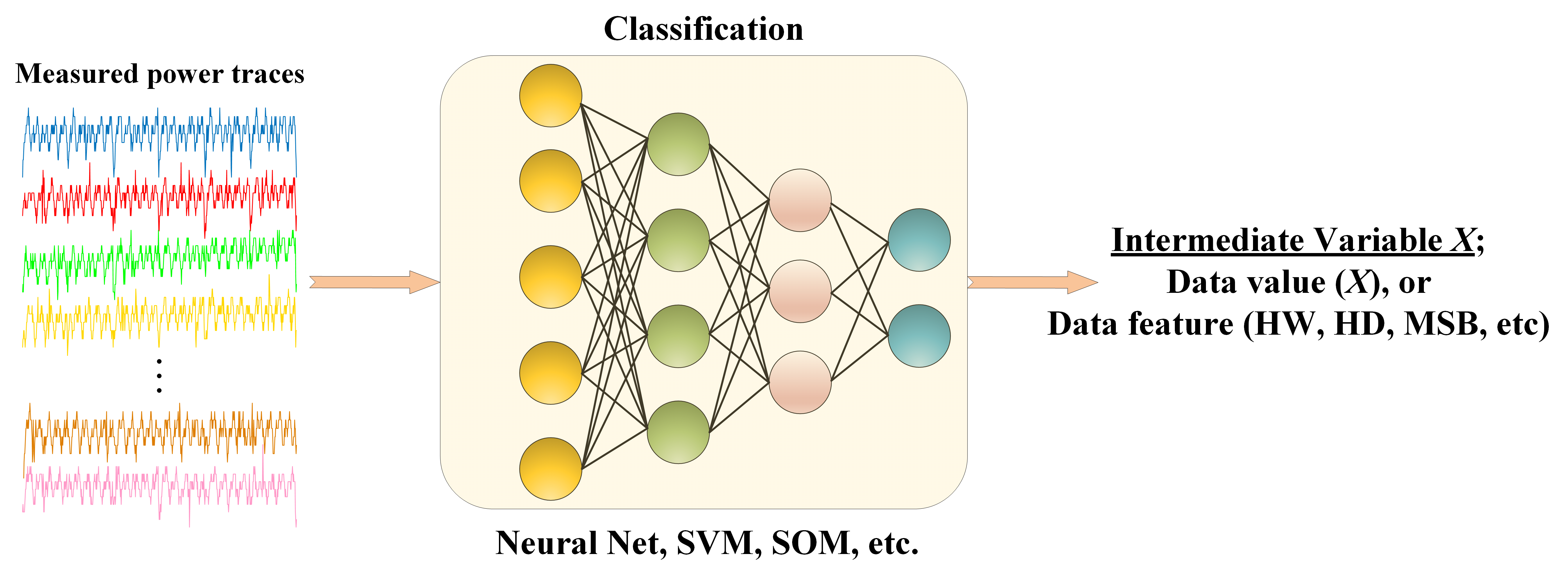}
	\vspace{-0.6cm}
	\caption{Conceptual description of supervised power analysis techniques.}
	\label{fig:supervised}
\end{figure}

\subsection{Supervised Learning Attacks}
Modern profiling-based techniques employ supervised learning to incorporate modeling of the leakage, extracting the proper statistics and decision metrics into a single algorithm, as shown in Fig.~\ref{fig:supervised}. In the terminology of supervised learning, the profiling phase of the attack corresponds to \textit{training} and the attack or exploitation phase is equivalent to \textit{inference} or \textit{test}. 

In supervised learning attacks, a set of power traces with known intermediate variables constitutes the training set. The label of a power trace, used in training, is the value of the corresponding intermediate variable. Supervised learning attacks have the flexibility to incorporate a leakage model if available. For example, if it is known that the power consumption is correlated with the HW of data, then the training labels would be the HW of the intermediate variable. Classical machine learning techniques use a leakage model for classification \cite{picek2017climbing, picek2017side}. Further, these techniques usually require a dimension reduction algorithm, such as principal component analysis (PCA), to select points of interest (POI) of power traces \cite{lerman2011side}.

Power analysis based on deep learning has been shown to be the most powerful profiling attack. The major advantages of deep learning include: 1) dimension reduction algorithms are not necessary to identify POIs; the neural networks can learn the most relevant features and decision metrics for best classification, 2) the data-dependent features can be identified even in misaligned traces, and 3) higher-order statistics required to attack masked implementations can be identified by neural networks. In \cite{maghrebi2016breaking}, convolutional neural networks (CNN), long short-term memory (LSTM) and a stacked auto-encoder are employed for feature extraction and  classification. The results show that deep learning techniques outperform classical machine learning algorithms such as SVM and RF. Multi-layer perceptron (MLP) neural networks have also been shown to outperform template attacks \cite{martinasek2015profiling}.

\subsection{Unsupervised Feature Extraction} \label{sec:autoecnoder}
Auto-encoders are the most prominent concepts of unsupervised learning in identifying data structure. Depending on the architecture of the encoder/decoder and the optimization goal, auto-encoders can be considered as generic nonlinear denoising filters and learning algorithms for estimating data distribution, identifying local manifold structure of data, dimension reduction and feature extraction. 

The basic concept of auto-encoders is simple. The input to the auto-encoder is $\hat{\mathrm{\mathbf{T}}}$ which is a corrupted, or noisy, version of the original data $\mathrm{\mathbf{T}}$. The encoder is a function $e():\mathbb{R}^N\to \mathbb{R}^D$ that maps the input data into an internal representation space and the decoder $d():\mathbb{R}^D\to \mathbb{R}^N$ maps the representation back into the input space to reconstruct the original data. The encoder and decoder functions are obtained by minimizing a loss function $\mathcal{L}(\mathrm{\mathbf{T}}, \hat{\mathrm{\mathbf{T}}})$ between the original and reconstructed data. In a denoising auto-encoder (DAE) \cite{vincent2008extracting}, the optimization goal is to minimize the mean of the loss function, i.e. $\mathrm{E[\mathcal{L}(\mathrm{\mathbf{T}}, \hat{\mathrm{\mathbf{T}}})]}$. The most popular loss functions are the squared error and cross-entropy.

Auto-encoders with a proper optimization goal can also be considered as manifold learning algorithms. A contractive auto-encoder (CAE) uses a regularization mechanism in the optimization problem to restrict the space of encoder/decoder functions. The optimization goal of CAE is
\begin{equation} \label{eq:cae}
    e,d = \argmin_{e,d} \mathrm{E}\Bigg[\mathcal{L}(\mathrm{\mathbf{T}}, \hat{\mathrm{\mathbf{T}}}) + \lambda \Big\|\frac{\partial e(\mathrm{\mathbf{T}})}{\partial \mathrm{\mathbf{T}}}\Big\|^2_F \Bigg]
\end{equation}
in which $\|.\|^2_F$ is the Frobenius norm and $\lambda$ is a design parameter. The first term in the above loss function is the reconstruction error and the second term is the regularization with $\lambda$ providing a trade-off between them.
If data is concentrated on a low-dimensional manifold, the CAE will learn a stochastic mapping from the input to the manifold. It is shown in \cite{vincent2008extracting} that DAE can also learn the data manifold when the dimension of the internal representation is constrained.


An alternative perspective to auto-encoders is an algorithm that extracts features with maximum mutual information with the data. Assume the encoder is a function $e()$ with parameters $W_e$, and the goal is to find features $\mathrm{\mathbf{f}}$ that have maximum mutual information with the original data $\mathrm{\mathbf{T}}$. The mutual information is
\begin{equation}
    I(\mathrm{\mathbf{T}};\mathrm{\mathbf{f}}) = H(\mathrm{\mathbf{T}}) - H(\mathrm{\mathbf{T}}|\mathrm{\mathbf{f}})
\end{equation}
Since the entropy of the data, i.e. $H(\mathrm{\mathbf{T}})$, is independent of the parameters $W_e$, maximizing the above mutual information reduces to minimizing the conditional entropy as
\begin{equation} \label{eq:condentropy}
    \hat{W}_e = \argmin_{W_e} H(\mathrm{\mathbf{T}}|\mathrm{\mathbf{f}})
\end{equation}
Using the definition of conditional entropy, we have
\begin{equation} \label{eq:minentropy}
    \hat{W}_e = \argmax_{W_e,\Tilde{p}} \mathrm{E}_{(\mathrm{\mathbf{T}},\mathrm{\mathbf{f}})}[ \log\Tilde{p}(\mathrm{\mathbf{T}}|\mathrm{\mathbf{f}}) ]
\end{equation}
in which the optimization problem is solved over all possible conditional probability density functions $\Tilde{p}(\mathrm{\mathbf{T}}|\mathrm{\mathbf{f}})$. However, when a particular structure of the decoder is chosen, the space of $\Tilde{p}(\mathrm{\mathbf{T}}|\mathrm{\mathbf{f}})$ is restricted and the criterion in (\ref{eq:minentropy}) provides a lower bound on the conditional entropy in (\ref{eq:condentropy}). 

The maximization problem (\ref{eq:minentropy}) can also be expressed in terms of the cross-entropy loss. Let $\hat{p}(\mathrm{\mathbf{T}}|\hat{\mathrm{\mathbf{T}}}; W_e,W_d)$ denote the conditional $\mathrm{\mathbf{T}}|\hat{\mathrm{\mathbf{T}}}$ in which the encoder and decoder parameters $W_e$ and $W_d$, respectively, are explicitly shown to emphasize that the density function is constrained by the structure of the encoder and the decoder. Note that $\mathrm{\mathbf{f}}=e(\hat{\mathrm{\mathbf{T}}})$ is a deterministic relation. The optimization problem in (\ref{eq:minentropy}) then amounts to
\begin{equation} \label{eq:xentropy}
    \hat{W}_e, \hat{W}_d = \argmin_{W_e,W_d}  \mathcal{H}\Big(p(\hat{\mathrm{\mathbf{T}}})\|\hat{p}(\mathrm{\mathbf{T}}|\hat{\mathrm{\mathbf{T}}}; W_e,W_d) \Big) 
\end{equation}
in which $\mathcal{H}(P\|Q)=\mathrm{E}_{P(X)}[-\log Q(X)]$ is the cross-entropy between densities $P$ and $Q$. Hence, to extract features that have maximum mutual information with data, the optimum parameters of the auto-encoder should minimize the cross-entropy between the distributions of the noisy measurements and the reconstructed data.

To gain further insights into the auto-encoder with the cross-entropy loss, we rewrite the conditional $\Tilde{p}(\mathrm{\mathbf{T}}|\mathrm{\mathbf{f}})$ in the maximization problem of (\ref{eq:minentropy}) in terms of $\hat{p}(\hat{\mathrm{\mathbf{T}}}|\mathrm{\mathbf{T}}; W_e,W_d)\hat{p}(\mathrm{\mathbf{T}}; W_e,W_d)$. Hence, the optimization problem reduces to
\begin{align}\label{eq:WeWd}
\begin{split}
    \hat{W}_e, \hat{W}_d &= \argmax_{W_e,W_d} \mathrm{E}_{(\mathrm{\mathbf{T}},\hat{\mathrm{\mathbf{T}}})} \Big[ \log\hat{p}(\hat{\mathrm{\mathbf{T}}}|\mathrm{\mathbf{T}}; W_e,W_d) \\
    &\hspace{1.2in} +\log p(\mathrm{\mathbf{T}}; W_e,W_d) \Big]\\
    &= \argmax_{W_e,W_d} \Big\{ \mathrm{E}_{(\mathrm{\mathbf{T}},\hat{\mathrm{\mathbf{T}}})} \Big[ \log\hat{p}(\hat{\mathrm{\mathbf{T}}}|\mathrm{\mathbf{T}}; W_e,W_d) \Big] - H(\mathrm{\mathbf{T}}) \Big\}
\end{split}
\end{align}
According to (\ref{eq:WeWd}), the optimization goal in a max information auto-encoder consists of two terms: 1) the expectation of the conditional log-likelihood $\hat{\mathrm{\mathbf{T}}}|\mathrm{\mathbf{T}}$ which should be maximized, and 2) the entropy of the reconstructed data that should be minimized. The second term can be considered as a regularization on the auto-encoder. Hence, the max information auto-encoder is a regularized maximum likelihood optimizer in which the regularization is minimizing the entropy of the reconstruction. 

Assume the corruption process is additive Gaussian noise, i.e. $\hat{\mathrm{\mathbf{T}}}=\mathrm{\mathbf{T}}+\mathrm{\mathbf{N}}$, in which $\mathrm{\mathbf{N}}\sim\mathcal{N}(0,\Sigma)$ and $\Sigma$ is the covariance of the noise. The conditional $\hat{\mathrm{\mathbf{T}}}|\mathrm{\mathbf{T}}$ is 
\begin{equation} \label{eq:Gauss}
    p(\hat{\mathrm{\mathbf{T}}}|\mathrm{\mathbf{T}}) = \frac{1}{\sqrt{(2\pi)^N|\Sigma|}} e^{-(\hat{\mathrm{\mathbf{T}}}-\mathrm{\mathbf{T}})^{T}\Sigma^{-1}(\hat{\mathrm{\mathbf{T}}}-\mathrm{\mathbf{T}})}
\end{equation}
Replacing this density into the optimization problem of (\ref{eq:WeWd}), we have
\begin{align} \label{eq:MSEloss}
\begin{split}
    & \hat{W}_e, \hat{W}_d = \\ 
    & \argmin_{W_e,W_d}  \Big\{ \mathrm{E}_{(\mathrm{\mathbf{T}},\hat{\mathrm{\mathbf{T}}})} \Big[ (\hat{\mathrm{\mathbf{T}}}-\mathrm{\mathbf{T}})^{T}\Sigma^{-1}(\hat{\mathrm{\mathbf{T}}}-\mathrm{\mathbf{T}}) \Big]+H(\mathrm{\mathbf{T}}) \Big\}
\end{split}
\end{align}
Assuming that the auto-encoder has converged to the optimal solution, the first term in (\ref{eq:MSEloss}) is simply the mean-square error (MSE) between the input and output of the auto-encoder while the second term is the regularization on the entropy of the output. 

In this paper, we assume a multivariate Gaussian distribution for the measurements, and we employ MSE as the loss function of the auto-encoder. Instead of the entropy regularization, we use the constraint on the dimension of the internal representation $\mathrm{\mathbf{f}}$ as the regularization. Since the distribution of the output, i.e. $\hat{p}(\mathrm{\mathbf{T}}; W_e,W_d)$, is constrained by the structure of the auto-encoder, lower dimension of the representation implies lower entropy of the output. Hence, under a Gaussian assumption for noise, the MSE auto-encoder can be considered as a sub-optimal solution for the max information auto-encoder.

According to the above discussion, the auto-encoder attenuates measurement noises which are stronger than data-dependent features and might result in incorrect estimates of the intermediate variables. Further, the auto-encoder encodes all data-related information of power traces into a low-dimensional internal representation. Hence, there is no need to identify points of interest in power traces. This is especially important with misaligned measurements where information about data is distributed over different samples. Finally, the auto-encoder provides a similarity model among measurements. Power traces with similar features have similar information content. This similarity might not be observable in raw measurements with noise and misalignment. The similarity model can be used to identify a leakage model.

\subsection{Leakage Modeling with Sensitivity Analysis} \label{sec:sensitivity}
Given the extracted features from power measurements, we identify which data features are encoded into the auto-encoder features using sensitivity analysis.
We recall the leakage model of (\ref{eq:leakmodel}) in which the individual terms constitute the \textit{data features}. We postulate that power traces with similar features are related to the intermediate values with similar data features.

For an $m$-bit intermediate variable $X$, the number of monomials $X^U$ in (\ref{eq:leakmodel}) is equal to $2^m-1$ with $U\in\mathbb{F}^m_2\backslash\{0\}$. Let $M_U():\mathbb{R}^D\to \mathbb{U}_d$ be an unbiased estimator of a monomial $X^U$ from power features in which $\mathbb{U}_d$ is the field of values for a monomial of degree $d=HW(U)$. The uncertainty in the estimate of a data feature $X^U$ can be measured by the information content of power traces about the data. 

A large amount of mutual information between the observation $\mathrm{\mathbf{f}}$ and a parameter $\theta=X^U$ means that the conditional entropy $H(\mathrm{\mathbf{f}}|\theta)$ is small, which implies that the conditional log-likelihood function $\log p(\mathrm{\mathbf{f}}|\theta)$ is concentrated. The first derivative of a concentrated log-likelihood function, called the \textit{score}, has large variations. The variance of the score is defined as Fisher information, i.e.
\begin{equation}
    \mathcal{I}(\theta) = \mathrm{E}_{\mathrm{\mathbf{f}}}\Big[\big(\frac{\partial}{\partial \theta} \log p(\mathrm{\mathbf{f}}|\theta)\big)^2 \Big]
\end{equation}

The inverse of Fisher information sets a lower bound, called the \textit{Cramer-Rao} bound, on the mean-square error of any unbiased estimator of the parameter $\theta$. Hence, the uncertainty in the estimate of a data feature $X^U$ is inversely proportional to the information content of measurements about data. If the power consumption is not correlated with a particular data feature $X^{U^*}$, i.e. small mutual information, the estimate of $X^{U^*}$ has a large uncertainty. This is the basis of \textit{sensitivity analysis} to identify those data features which are highly correlated with the power traces.

Using a Taylor series expansion for an unbiased estimator $\hat{\theta}=M_U(\mathrm{\mathbf{f}})$ around its mean, i.e. $\theta$, we have
\begin{equation} \label{eq:taylor}
    \hat{\theta} \approx \theta + \nabla M_U^T\cdot \triangle \mathrm{\mathbf{f}}
\end{equation}
in which $\nabla M_U$ is the gradient of the estimator. The Cramer-Rao bound on the variance of the estimate is the inverse of the Fisher information as
\begin{equation} \label{eq:crb}
    \mathrm{Var}(\hat{\theta}) \ge \mathcal{I}^{-1}(\theta)
\end{equation}
Using equation (\ref{eq:taylor}) in (\ref{eq:crb}), we can express the following bound on the Hessian matrix of the estimator, i.e. $\nabla M_U\nabla M_U^T$,
\begin{equation}
    \mathrm{E}_{\mathrm{\mathbf{f}}}\Big[\triangle \mathrm{\mathbf{f}}^T\nabla M_U\nabla M_U^T\triangle \mathrm{\mathbf{f}} \Big] \ge \mathcal{I}^{-1}(\theta)
\end{equation}
This relation implies that the curvature of the estimator around the mean is lower-bounded by the inverse of the Fisher information. 
By adding a small perturbation to the function $M_U()$, the estimate of the parameter will change to $\Tilde{\theta} \approx \theta + \nabla\Tilde{M}_U^T\cdot \triangle \mathrm{\mathbf{f}}$
in which $\nabla\Tilde{M}_U=\nabla M_U+\delta$ and $\delta$ is a small perturbation with a magnitude much smaller than the gradient $\nabla M_U$. Assuming $\mathrm{\mathbf{b}}^T\nabla\Tilde{M}_U\nabla M_U^T\mathrm{\mathbf{b}}\approx \mathrm{\mathbf{b}}^T\nabla M_U\nabla M_U^T\mathrm{\mathbf{b}}$ for any vector $\mathrm{\mathbf{b}}$, the variance of the change in the estimate as a result of perturbations is then
\begin{equation}
    \mathrm{E}_{\mathrm{\mathbf{f}}} \Big[(\Tilde{\theta}-\hat{\theta})^2 \Big] \ge \mathcal{I}^{-1}(\theta)
\end{equation}

The above analysis shows that if the observations $\mathrm{\mathbf{f}}$ have a large information content about a parameter $\theta$, the sensitivity of an estimate $\hat{\theta}=M_U(\mathrm{\mathbf{f}})$ to slight perturbations in the estimator function is small. This is consistent with the analysis of shrinkage amount of coefficients in a stochastic leakage model as discussed in \cite{wang2017ridge}.

\begin{figure*}[htbp]
	\centering
	\begin{multicols}{2}
		\includegraphics[width=1\textwidth]{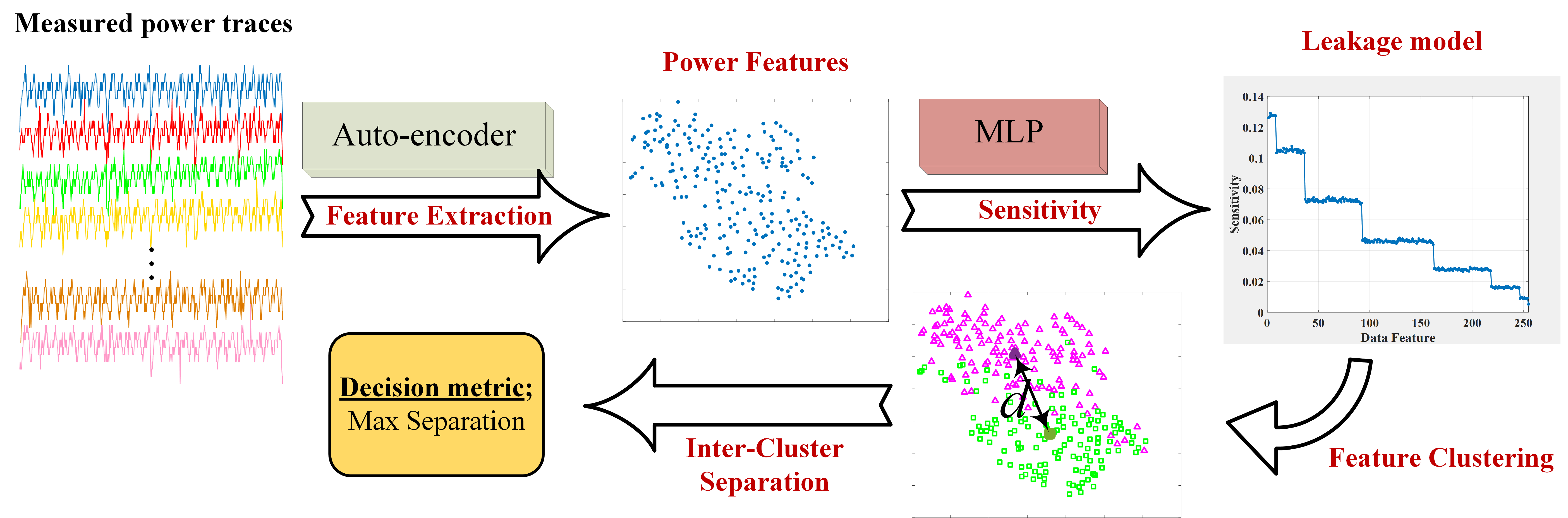}
	\end{multicols}
	\vspace{-0.6cm}
	\caption{Conceptual description of the proposed unsupervised learning technique for power analysis (SCAUL).}
	\label{fig:scaul}
\end{figure*}

\section{SCAUL Methodology} \label{sec:method}
The overall procedure of the proposed side-channel analysis with unsupervised learning (SCAUL) approach is shown in Fig. \ref{fig:scaul}. The steps of SCAUL are explained below.
\begin{enumerate}
    \item An LSTM auto-encoder extracts features from the measured power traces. As the auto-encoder identifies the data-dependent features in an unsupervised approach, the traces may correspond to any intermediate variable. 
    \item A multi-layer perceptron (MLP) neural network is trained to estimate the bits of the intermediate variable from the power features for all key candidates.
    \item Following a sensitivity analysis, a slight perturbation is added to the weights of the trained MLP neural network. The variation of data features at the output of the MLP as a result of the perturbation is measured. Data features with lowest variation constitute the leakage model.
    \item The power features, extracted by the auto-encoder, are clustered based on the identified leakage model. The correct key exhibits the highest inter-cluster difference.
\end{enumerate}
The details on the structure of the neural networks in the SCAUL technique are explained in the following sections.

\subsection{LSTM Auto-encoder}
Recurrent neural networks (RNN) are popular for learning temporal models of time series in applications such as natural language processing (NLP) \cite{yin2017comparative}, speech recognition and acoustic modeling \cite{sak2014long}. A Long short-term memory (LSTM) neural network is a special type of RNN that can learn both local (short-term) and long-term temporal dependence of a signal. Convolutional neural networks (CNNs) have also been employed for processing time series. While CNNs are powerful in learning position invariance features, LSTM networks are stronger in learning temporal models \cite{sainath2015convolutional}.

\begin{figure}[t!]
	\centering
	\includegraphics[width=0.35\textwidth]{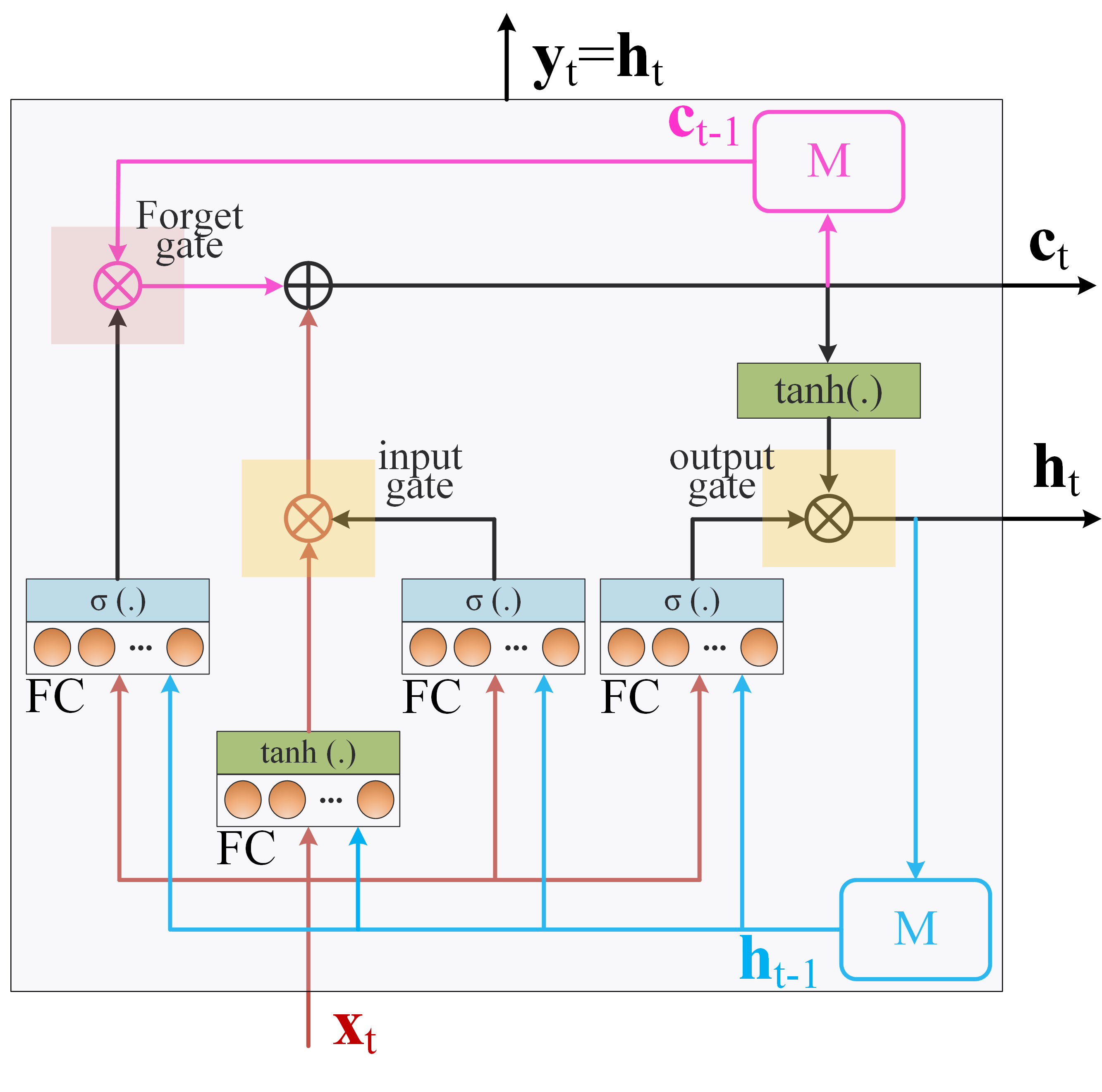}
	\vspace{-0.6cm}
	\caption{The basic cell in long short-term memory (LSTM) neural network.}
	\label{fig:lstmcell}
\end{figure}

The basic cell of an LSTM neural network is shown in Fig. \ref{fig:lstmcell}. It consists of internal states $\mathrm{\mathbf{c}}$ and $\mathrm{\mathbf{h}}$, the latter of which is the output of the cell at every time instance. The input to the cell is processed by a fully-connected (FC) network with $tanh$ activation. The internal state at each time instant is a weighted sum of the previous state and the processed input. The weighting process, called \textit{gate}, controls the memory of the cell. The \textit{forget gate} determines how much information of previous states is preserved at the current time instant and the \textit{input gate} controls the amount of input activation to be added to the state. The output of the cell at time $t$, i.e. $\mathrm{\mathbf{h}}_t$, is calculated by applying $tanh$ activation to the internal state $\mathrm{\mathbf{c}}_t$ followed by the \textit{output gate}. This gate controls how much of the internal state activation flows to the output. The weights used in the above three gates are determined by three FC units with sigmoid activation. The input to these FC units is the current input to the cell, i.e. $\mathrm{\mathbf{x}}_t$, and the previous output, i.e. $\mathrm{\mathbf{h}}_{t-1}$.

\begin{figure*}[htbp]
	\centering 
	\begin{multicols}{2}
		\includegraphics[width=0.9\textwidth]{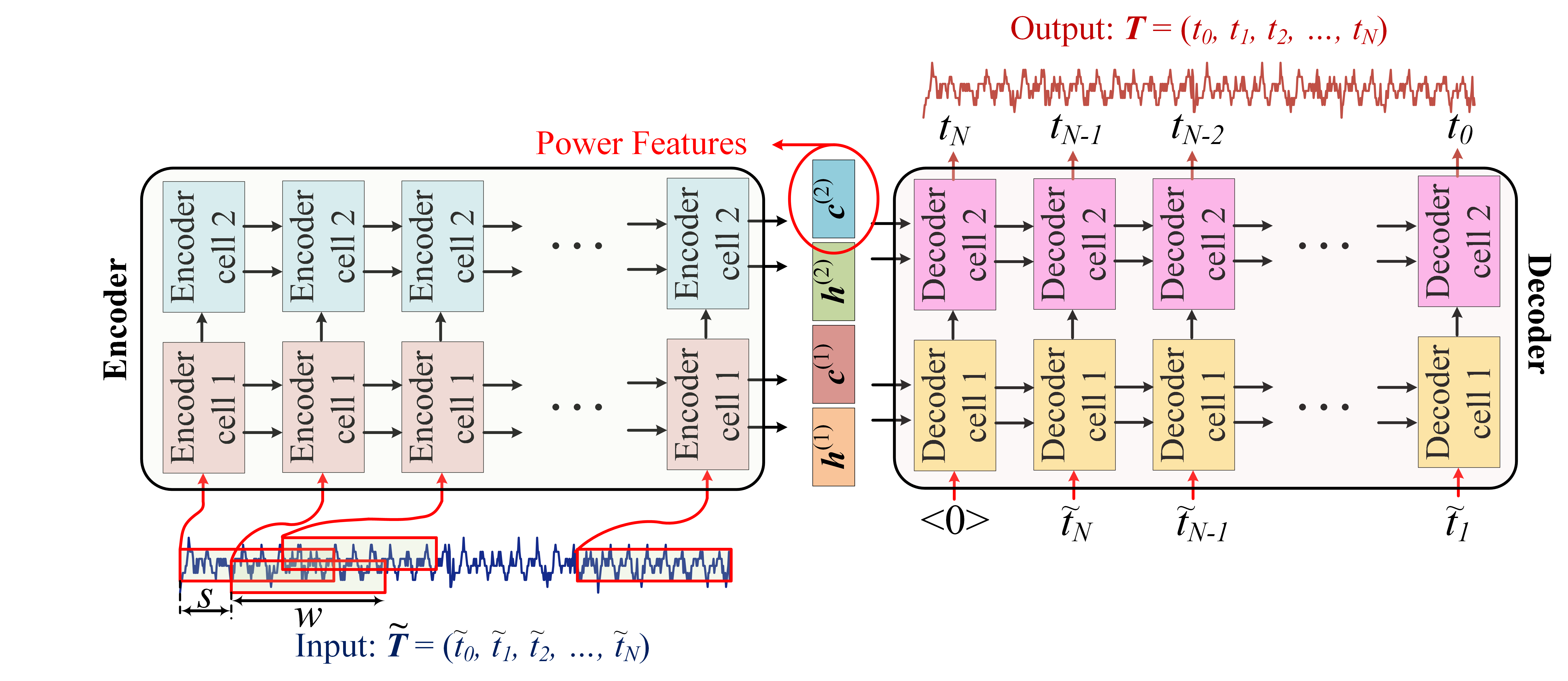}
	\end{multicols}
	\vspace{-0.8cm}
	\caption{Proposed LSTM auto-encoder for extracting features of power traces, with sliding window processing of input power traces. and $\mathrm{\mathbf{c}}$ state of top encoder cell selected as power feature.}
	\label{fig:lstm_auto}
\end{figure*}

The proposed LSTM auto-encoder for feature extraction is shown in Fig. \ref{fig:lstm_auto}. The encoder and decoder are LSTM networks with two layers shown in a \textit{time-unrolled} representation; each of the encoder and decoder consists of a stack of two LSTM cells, as in Fig.~\ref{fig:lstmcell}, however, the processing steps of the cell are unrolled through time with the corresponding input to the cell shown explicitly at every time instant.

The inputs to the encoder are the samples of the power traces provided by a sliding window of length $w$ and a stride of $s$; i.e., at every time instant, the input is a vector of $w$ consecutive samples of the trace and the window is offset by $s$ samples relative to the previous time instant. This is similar to a convolutional layer in CNNs. The first input to the decoder is zero while the following inputs are the individual samples of the power trace in the reverse order. The outputs of the decoder are the samples of the filtered power traces in the reverse order. 

The loss function for optimizing the parameters of the auto-encoder is the MSE between the input of the encoder and output of the decoder. A constraint on the internal state of the encoder/decoder cells is considered as a regularization. After training the auto-encoder with all available power traces, the internal state $\mathrm{\mathbf{c}}$ of the top-layer encoder cell is chosen as the features of power traces. We point out that the $\mathrm{\mathbf{c}}$ state contains most of the information about the data. According to Fig. \ref{fig:lstmcell}, the $\mathrm{\mathbf{h}}$ state is derived from the $\mathrm{\mathbf{c}}$ state. Further, the $\mathrm{\mathbf{c}}$ state of the top layer also contains information processed by the preceding layers. 

\subsection{Leakage Modeling with MLP}
After extracting data-dependent features from power traces with the auto-encoder, we estimate the intermediate variable from the features with a multi-layer perceptron (MLP) neural network. We train an MLP network for every key candidate in which the input is the power features, and the output is the bits of the intermediate variable calculated for the key candidate. 

\begin{figure}[t!]
	\centering
	\includegraphics[width=0.4\textwidth]{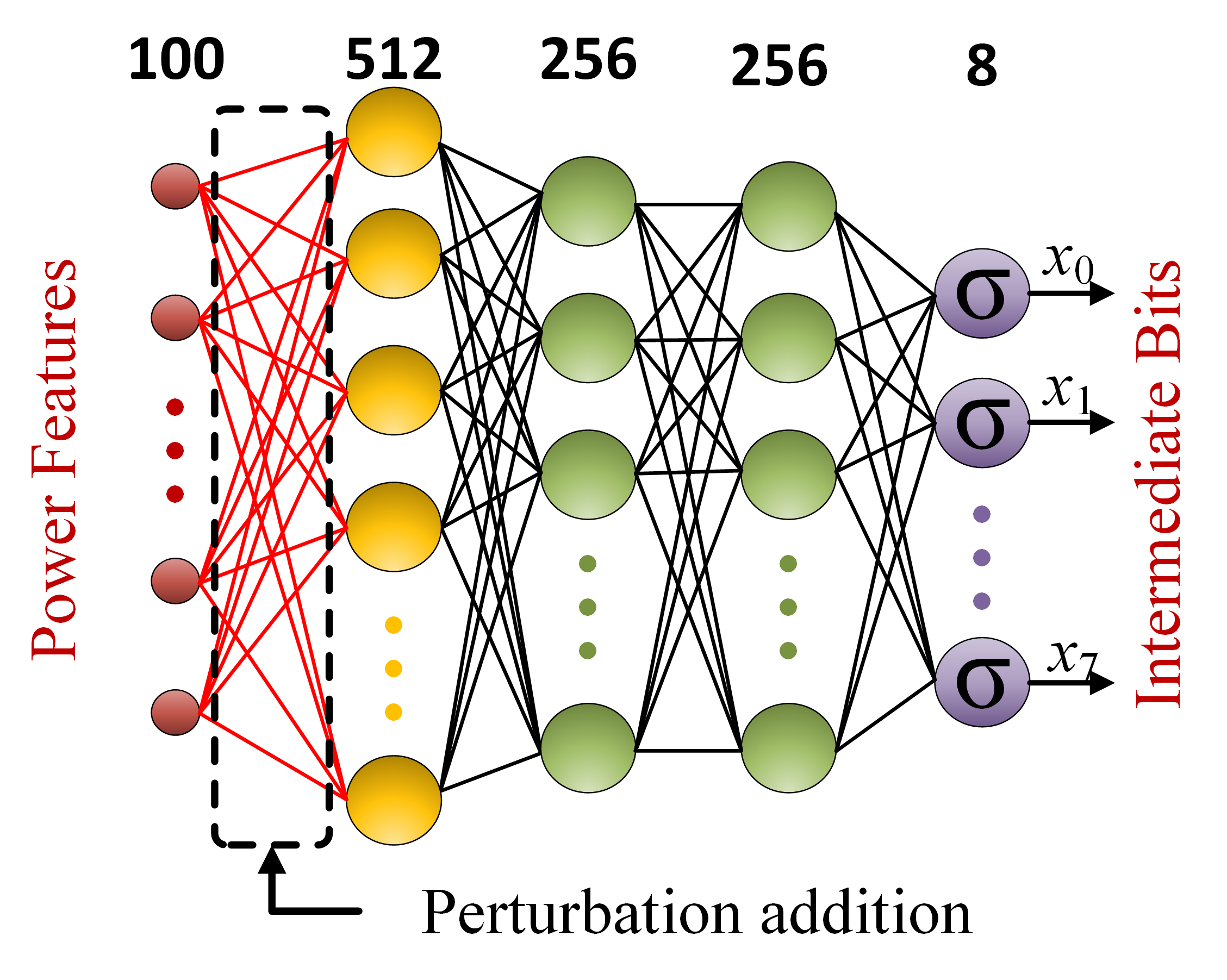}
	\vspace{-0.5cm}
	\caption{Multi-layer perceptron (MLP) neural network for estimating bits of intermediate variable from power features and sensitivity analysis.}
	\label{fig:mlp}
\end{figure}

The architecture of the MLP used in this work is shown in Fig. \ref{fig:mlp}. It consists of three hidden layers with ReLU activation. The output layer is a set of $m$ neurons, corresponding to $m$ bits of the intermediate variable, with sigmoid activation. The number of neurons at each layer, used in our experiments for power analysis of AES, is shown in the figure. The size of power features, that is equal to the size of the $\mathrm{\mathbf{c}}$ state of the LSTM auto-encoder, is 100, and the intermediate variable is one byte (8 bits) of the AES state at the output of an S-box. 

To facilitate training of the MLP, the input power features are normalized over all measurements. Let $\mathrm{\mathbf{c}}_j^{(2)}, j=1,2,\cdots,S$ denote the power features, extracted at the top-layer encoder cell of the auto-encoder in Fig. \ref{fig:lstm_auto}, corresponding to $S$ power measurements. The input to the MLP is then
\begin{equation} \label{eq:pfeatures}
    \Tilde{\mathrm{\mathbf{f}}}_j = \frac{\mathrm{\mathbf{c}}_j^{(2)}-\min\limits_{j}(\mathrm{\mathbf{c}}_j^{(2)})}{\max\limits_{j}(\mathrm{\mathbf{c}}_j^{(2)})-\min\limits_{j}(\mathrm{\mathbf{c}}_j^{(2)})}
\end{equation}
in which the normalization is carried out element-wise. Hence, the inputs to the MLP are within $[0,1]$. The outputs are the bits of the intermediate variable corresponding to the $j$-th measurement denoted by $\mathrm{\mathbf{x}}_j=(x_{j,0},x_{j,1},\cdots,x_{j,m-1})$ with values in $\{0, 1\}$. 

After training the MLP with the normalized power features in (\ref{eq:pfeatures}) and the corresponding intermediate variable $\mathrm{\mathbf{x}}_j$, a small perturbation is added to the weights of the MLP. The perturbation is added at the first layer as shown in Fig.~\ref{fig:mlp}. Let $\mathbf{W}_{0,1}$ denote the weights of the trained MLP from input to the first hidden layer. The perturbed network has the weights $\Tilde{\mathbf{W}}_{0,1} = \mathbf{W}_{0,1} + \mathbf{\delta}$ in which $\mathbf{\delta}$ is a small constant. The estimated intermediate variable at the output of the perturbed network is $\Tilde{\mathrm{\mathbf{x}}}_j=(\Tilde{x}_{j,0},\Tilde{x}_{j,1},\cdots,\Tilde{x}_{j,m-1})$. We calculate the variation of the perturbation in a monomial $\Tilde{X}_j^U=\prod_{i=0}^{m-1}\Tilde{x}_{j,i}\cdot u_i$ for $U\in \mathbb{F}_2^m\backslash \{0\}$ as
\begin{equation} \label{eq:perturb}
    \Delta_U = \mathrm{E}_j \Big[\big|\Tilde{X}_j^U - X_j^U \big|\Big]
\end{equation}
in which the expectation is over all measurements. 

According to the analysis of Section \ref{sec:sensitivity}, the lower variation of (\ref{eq:perturb}) implies that the data feature has a more significant contribution to the power consumption. Hence, in the leakage model of (\ref{eq:leakmodel}), we set
\begin{equation}
    \hat{\alpha}_U = 1-\frac{\Delta_U}{\max\limits_U \Delta_U}
\end{equation}
We cluster the coefficients based on the variations of (\ref{eq:perturb}), and select the coefficients in the cluster with smallest variations as the leakage model. This is similar to the constraint on the degree of the model in the ridge regression technique of \cite{wang2017ridge}. 

The power features are divided into two clusters based on the leakage model, obtained in an unsupervised approach. The correct key exhibits the highest inter-cluster difference. This is the fundamental property exploited in power analysis. Inter-cluster difference implies data-dependency of power features.

\section{Case Study on AES} \label{sec:AES}
Advanced encryption standard (AES) is a worldwide standard for secret-key cryptography. Several block ciphers have also adopted structures similar to AES. In this section, we demonstrate the SCAUL attack on an FPGA implementation of AES. The principles of the attack are the same for any cipher with a key-dependent operation in which the input or output is known, and its power consumption is dependent on the processed data.

The secret state of AES consists of 128 bits arranged in $4\times 4$ bytes. The cipher operations are carried out in 10 rounds with a composition of four transformations: \textit{AddRoundKey}, \textit{SubBytes}, \textit{ShiftRows} and \textit{MixColumns}. At the beginning, the plaintext is loaded into the state and the secret key is added. The next operation is S-box, or \textit{SubBytes}, which is a nonlinear transformation operating on individual bytes of the state. Next, \textit{ShiftRows} and \textit{MixColumns} operations follow to complete one round. 

The target of a typical power attack on AES is the S-box operation in round 1. By denoting $i$-th byte of the plaintext and the secret key as $P_i$ and $K_i$, respectively, the key-dependent cipher operation under attack is $X_i=S(P_i\oplus K_i)$ in which $S()$ is the S-box function. The intermediate variable is $X_i$ which is unknown but correlated with the power consumption during the operation of $S()$. Using the power measurements, the intermediate variable can be estimated. Hence, the input to the S-box, i.e. $P_i\oplus K_i$, can be calculated using the inverse S-box operation. Given the plaintext byte $P_i$, the corresponding byte of the secret key, i.e. $K_i$, will be recovered. 

\begin{figure}[t!]
	\centering
	\includegraphics[width=0.5\textwidth]{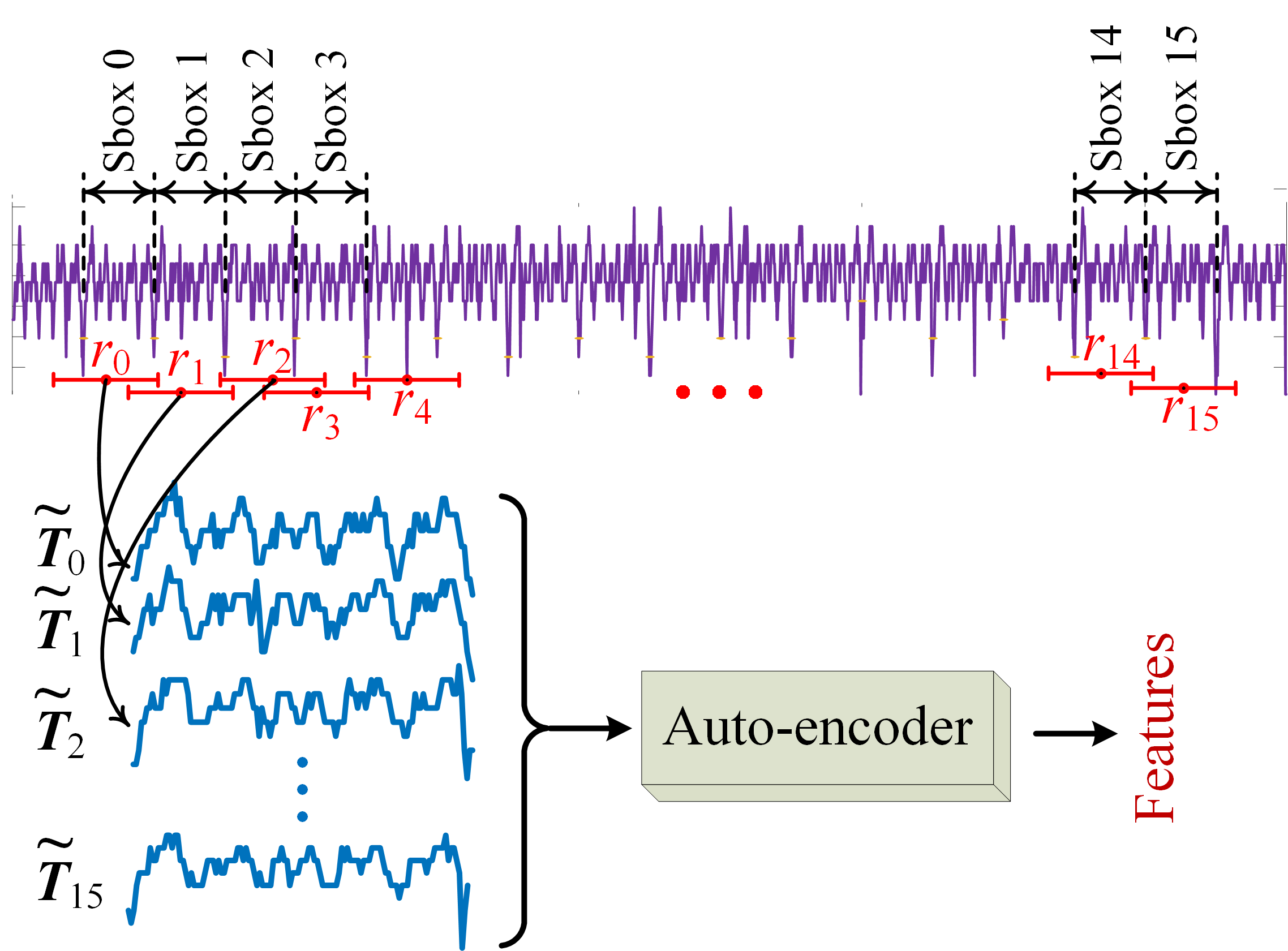}
	\vspace{-0.6cm}
	\caption{Horizontal processing of power measurements during round 1 of AES by the auto-encoder.}
	\label{fig:horizon}
\end{figure}

In our experiments, we use a lightweight implementation of AES (236 slices) on Artix-7 FPGA. The S-box function is implemented with a look-up table (LUT). At every clock cycle, the S-box is applied on one byte of the state. Hence, the power trace of round 1 corresponds to 16 S-box operations, as shown in Fig. \ref{fig:horizon}. The power traces corresponding to S-box operations are selected using measurement windows of length $l$ at positions $r_i, i=0,1,\cdots, 15$. The length of the windows are chosen based on the uncertainty in the timing of the measurements. If there is an uncertainty of $\Delta l$ between the measurements and the clock signal of the hardware, the length $l$ must be at least $l_c+\Delta l$, in which $l_c$ is the length of a clock cycle, so that the power trace include all power samples of the corresponding S-box operation.

Processing of power traces as in Fig. \ref{fig:horizon} is similar to the \textit{horizontal} attacks of \cite{aysu2018horizontal, chen2015horizontal} in which similar patterns of power consumption through time, corresponding to the same key subset, are analyzed to recover the key. However, in the SCAUL attack on AES, the power traces correspond to different key subsets. The main mechanism enabling such horizontal processing of power traces in AES is the feature extraction via auto-encoders. The auto-encoder can identify data-dependent features irrespective of the value of the intermediate variable. Using all power traces through time improves the accuracy of feature detection, hence, significantly reduces the required number of power measurements to recover the key, as shown in the next section.

\begin{figure}[t!]
	\centering
	\includegraphics[width=0.4\textwidth]{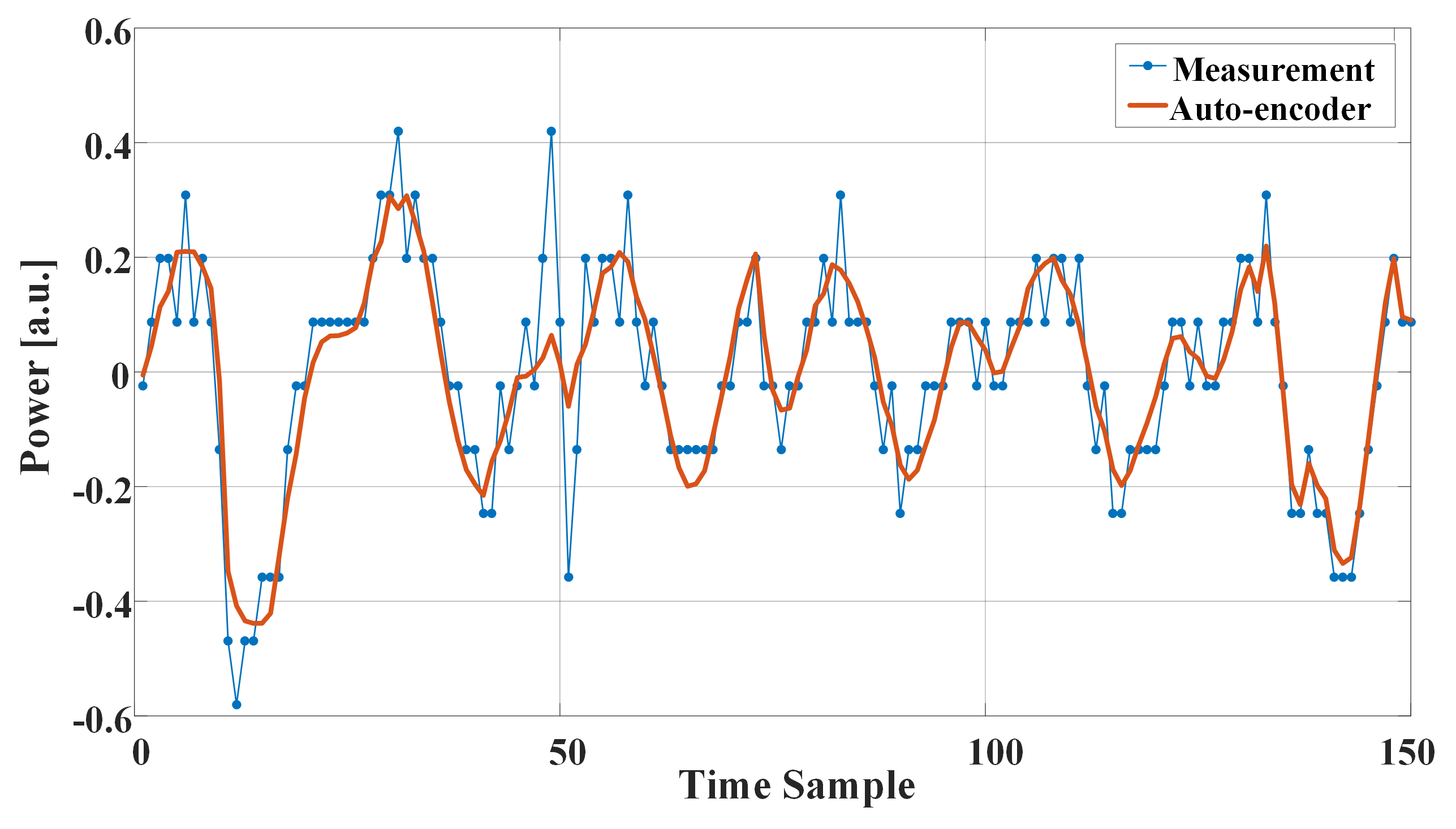}
	\vspace{-0.4cm}
	\caption{Power trace measured during operation of a S-box and filtered by LSTM auto-encoder.}
	\label{fig:filter}
\end{figure}

\begin{figure}[t!]
	\centering
	\includegraphics[width=0.4\textwidth]{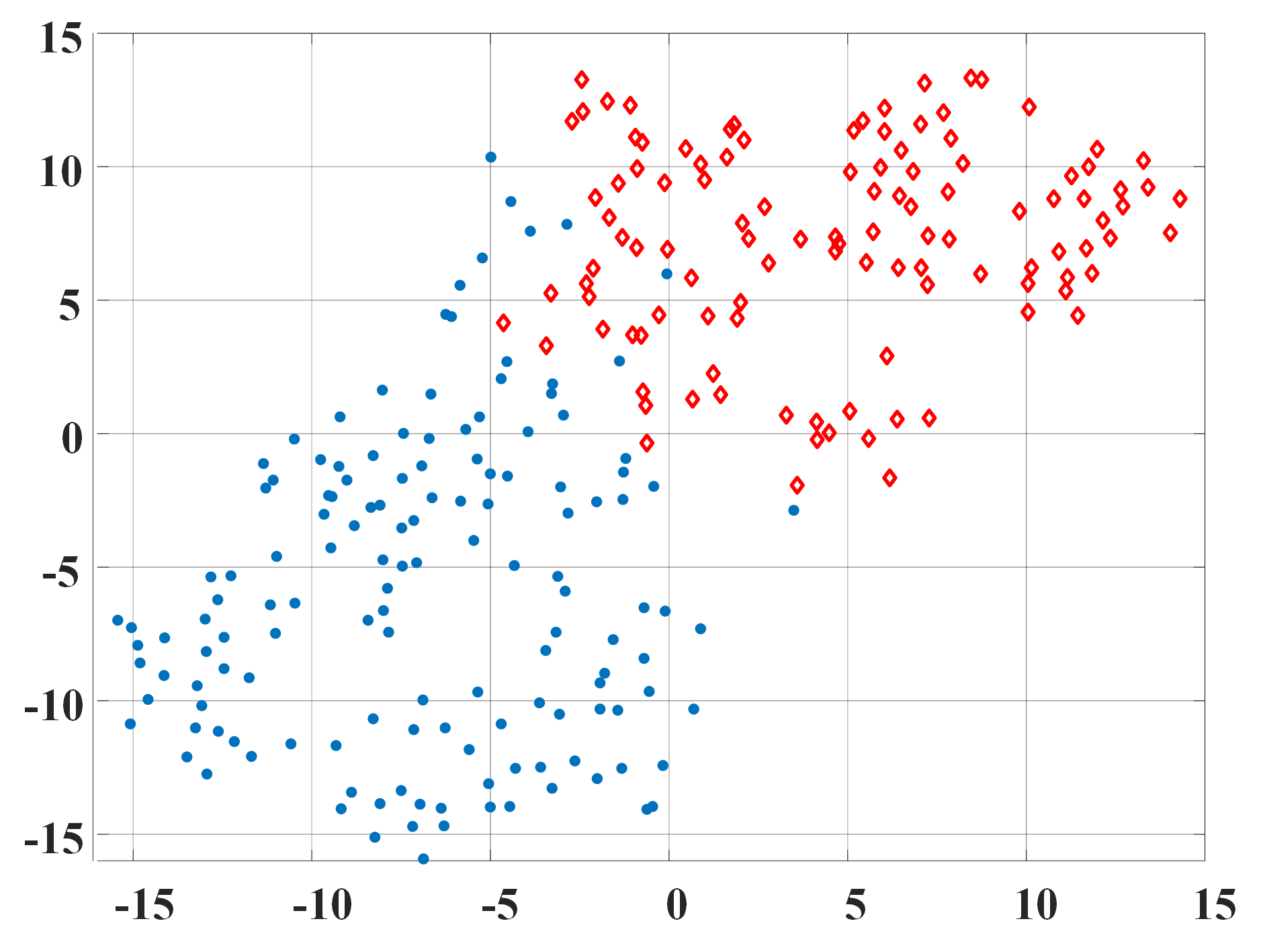}
	\vspace{-0.4cm}
	\caption{Extracted features from power consumption of S-box operations in AES plotted in 2 dimensions using t-SNE algorithm.}
	\label{fig:features}
\end{figure}

The filtering effect of the LSTM auto-encoder on power measurements is shown in Fig. \ref{fig:filter}. It is observed that while the underlying patterns of the power consumption are preserved at the output of the auto-encoder, strong noisy samples are filtered. The auto-encoder learns the patterns that repeat in most traces and filters out instantaneous variations that have low mutual information with the measurements. The extracted features from the power traces with an LSTM auto-encoder with 100 neurons in its FC components are also shown Fig. \ref{fig:features}. 

Since the LSTM auto-encoder has 100 neurons in the FC components, the extracted features also have a dimension of 100. The features are shown in the 2-dimensional plot of Fig. \ref{fig:features} using t-SNE algorithm \cite{maaten2008visualizing}. Each point in the plot represents the mean of all features corresponding to the same intermediate variable. The non-uniform distance between the points reflects data-dependency; the intermediate values with similar power features result in \textit{similar} power consumption. However, this similarity is not necessarily on individual samples of power traces. Instead, the data-dependent features of the traces, which might happen at different time samples, are similar. 

\begin{figure}[t!]
	\centering
	\includegraphics[width=0.4\textwidth]{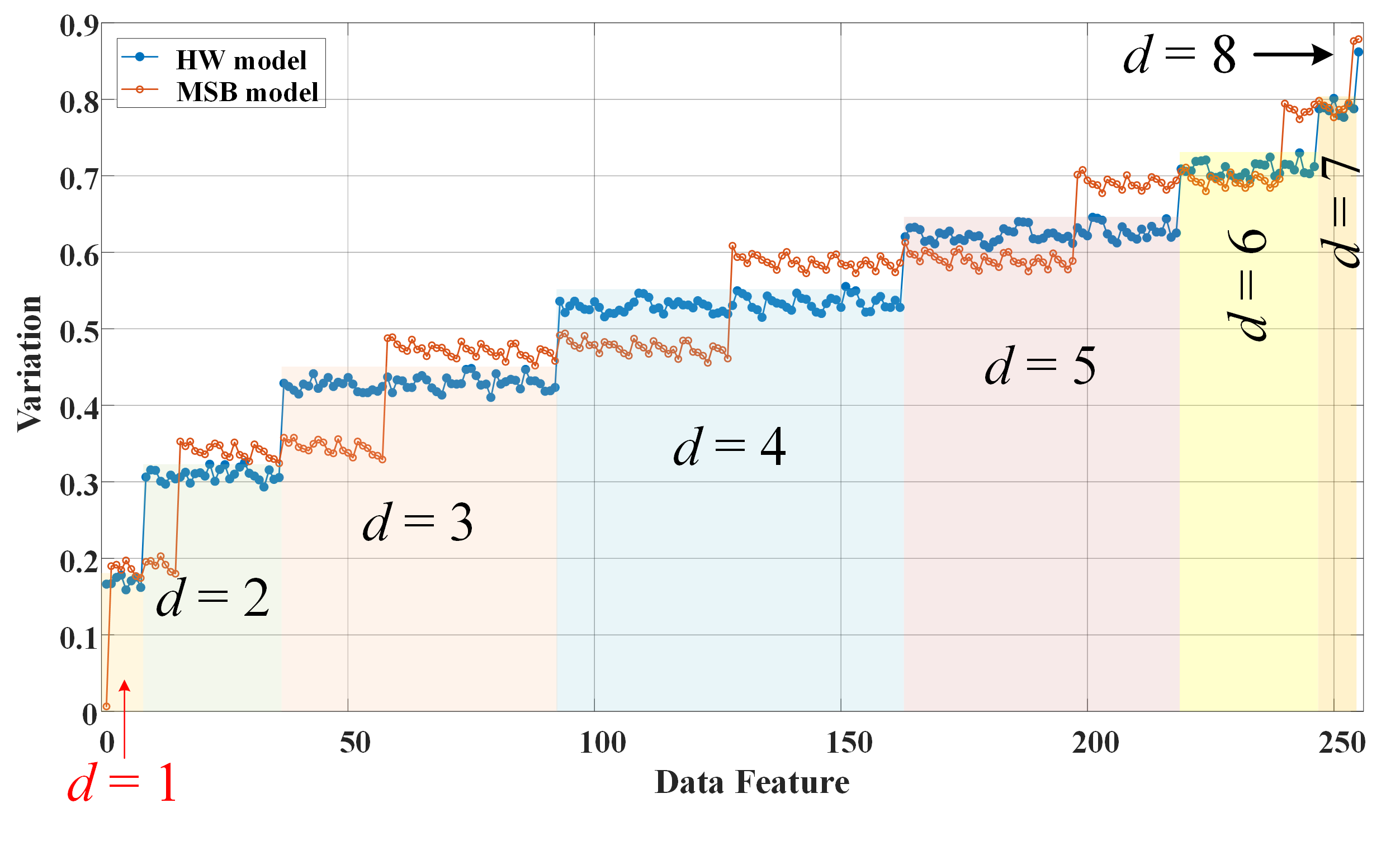}
	\vspace{-0.4cm}
	\caption{Variation of data features calculated with sensitivity analysis on MLP neural network.}
	\label{fig:dmodel}
\end{figure}

The power features extracted by the auto-encoder are mapped to the bits of intermediate variable using the MLP network of Fig. \ref{fig:mlp}. According to the sensitivity analysis of Section \ref{sec:sensitivity}, the coefficients of the leakage model are estimated. The variation of the monomials in (\ref{eq:leakmodel}) as a result of small perturbation on the weights of the trained MLP for the correct key candidate, calculated according to (\ref{eq:perturb}), is shown in Fig. \ref{fig:dmodel}. The curve labeled with ``HW model" is obtained for actual power measurements on FPGA. It is observed that the variation of monomials corresponding to degree $d=1$ are similar and lower than higher degree terms. This implies a HW leakage model; individual bits have the largest contribution to the power consumption with the same weights. This model can be verified using a DPA attack with an HW model, as shown in the next section.

To further verify the capability of the sensitivity analysis with MLP, we conduct a hypothetical experiment as follows. We group the power features into two clusters as shown in Fig. \ref{fig:features}. Then, we assign the values of the intermediate variable with the most significant bit (MSB) of 1 to one cluster and the values with MSB of 0 to the other cluster. We train an MLP with the power features and this hypothetical intermediate variable. The variation of the data features as a result of perturbation on the MLP weights is also shown in Fig. \ref{fig:dmodel} labeled with ``MSB model". It is observed that the variation corresponding to MSB (the first data feature) has the lowest variation. It implies that the power consumption is correlated with the MSB of the intermediate variable.

\section{Experimental Results} \label{sec:results}
We demonstrate the SCAUL attack on an FPGA implementation of AES using the Flexible Open-source workBench fOr Side-channel analysis (FOBOS) \cite{fobos}. The FOBOS instance uses a NewAE CW305 Artix-7 FPGA target for the AES implementation, and Digilent Nexys 7 as the control board for synchronization with a host PC and target FPGA. We measure the power consumption of the target FPGA during encryption of multiple random plaintexts with 125 samples per clock cycle. 

In all our experiments, we use the LSTM auto-encoder of Fig. \ref{fig:lstm_auto} with 100 neurons in the FC components of both encoder and decoder cells. The sliding window at the input of the encoder has a length of 10 samples and a stride of 2. The LSTM auto-encoder and MLP neural network of Fig. \ref{fig:mlp}, for sensitivity analysis, are implemented in Tensorflow. The Adaptive moment estimation (Adam) algorithm is used for training all neural networks. 

On a PC with Intel Core-i7 CPU, 16 GB RAM, and Nvidia GeForce GTX 1080 GPU, the LSTM auto-encoder takes around 20 minutes to train with more than 20K traces. Training of the MLP network with sensitivity analysis requires around 25 seconds for a key candidate and around 1.8 hours for all 256 possible values of the key candidate. The auto-encoder and the extracted leakage model are re-used for all bytes of the entire secret key. Including clustering, the overall time for a SCAUL attack would take around 2.5 hours for recovering the entire 128-bit key of AES.

\subsection{Power Analysis with Leakage Model}
In the first experiment, we conduct a model-based power attack using the HW leakage model as a basis for comparing the performance of the proposed unsupervised learning approach in recovering the leakage model and the correct key. We employ a DPA attack using both the individual power samples, as in classical techniques, and the power features extracted by the LSTM auto-encoder. The latter reveals the capability of the auto-encoder in extracting data-dependent features.

\begin{figure}[t!]
	\centering
	\includegraphics[width=0.4\textwidth]{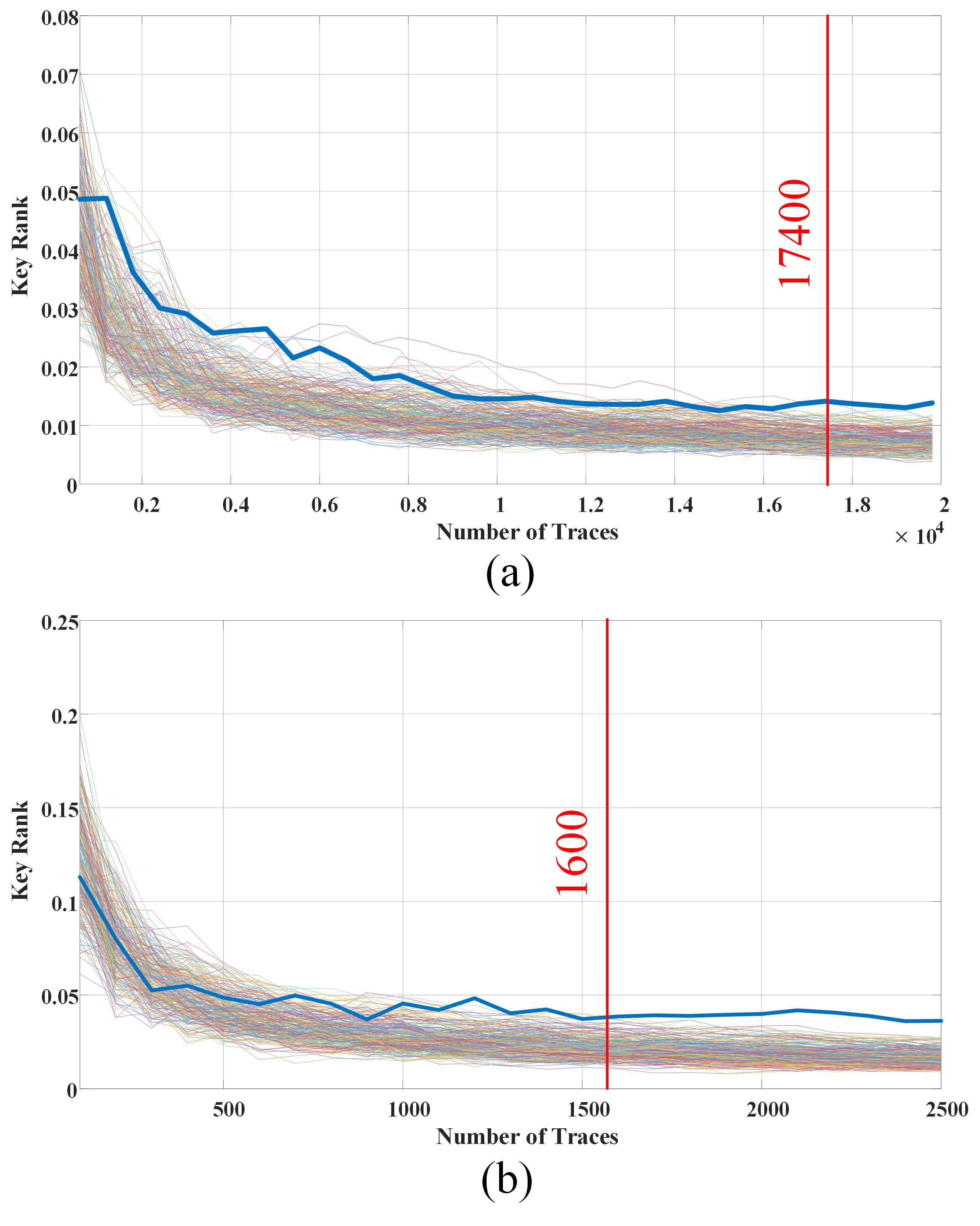}
	\vspace{-0.4cm}
	\caption{Rank of key candidates versus number of traces in a DPA attack with HW leakage model; (a) maximum difference of power samples, (b) maximum difference of power features.}
	\label{fig:dpa}
\end{figure}

The results of a classical DPA attack with the HW model are shown in Fig. \ref{fig:dpa} (a). For every key candidate the intermediate variable $X$ is calculated based on which the power traces are grouped into two clusters $C_0$ and $C_1$; power traces corresponding to the intermediate variable $X$ with $HW(X)<4$ are in cluster $C_0$ and those with $HW(X)>4$ belong to $C_1$. The mean trace of each cluster is obtained by averaging all traces in the cluster. The absolute values of the difference between samples of the mean traces in two clusters are calculated. The rank of a key candidate is the maximum absolute difference. 

It is observed in Fig. \ref{fig:dpa} (a) that with at least 17400 power traces, corresponding to the power measurement during the encryption of 17400 random plaintexts, the rank of the correct key is always larger than all incorrect key candidates in a DPA attack. Considering the fact that the attack is successful with the HW model also justifies the results of Fig. \ref{fig:dmodel} in which the sensitivity analysis with the MLP suggests an HW leakage model.

To verify the effect of the auto-encoder, we repeat a DPA attack similar to the above attack by replacing the raw power traces with the power features of (\ref{eq:pfeatures}) extracted by the auto-encoder. The rank of key candidates versus the number of power measurements (encryptions) is shown in Fig. \ref{fig:dpa} (b). It is observed that using the power features, only 1600 measurements are sufficient to identify the correct key, i.e., an improvement of more than $10\times$ in attack efficiency compared to classical DPA.

The significant effect of the auto-encoder in improving the performance of the DPA attack implies that the auto-encoder extracts the most relevant features of the power traces that depend on the processed data. The noisy samples of the power measurements add constructively in a DPA attack for some key candidates which results in large inter-cluster difference and hinders detection of the correct key. However, the auto-encoder has the ability to identify data-dependent features even in the presence of noise.

\subsection{SCAUL Attack}
\begin{figure}[t!]
	\centering
	\includegraphics[width=0.4\textwidth]{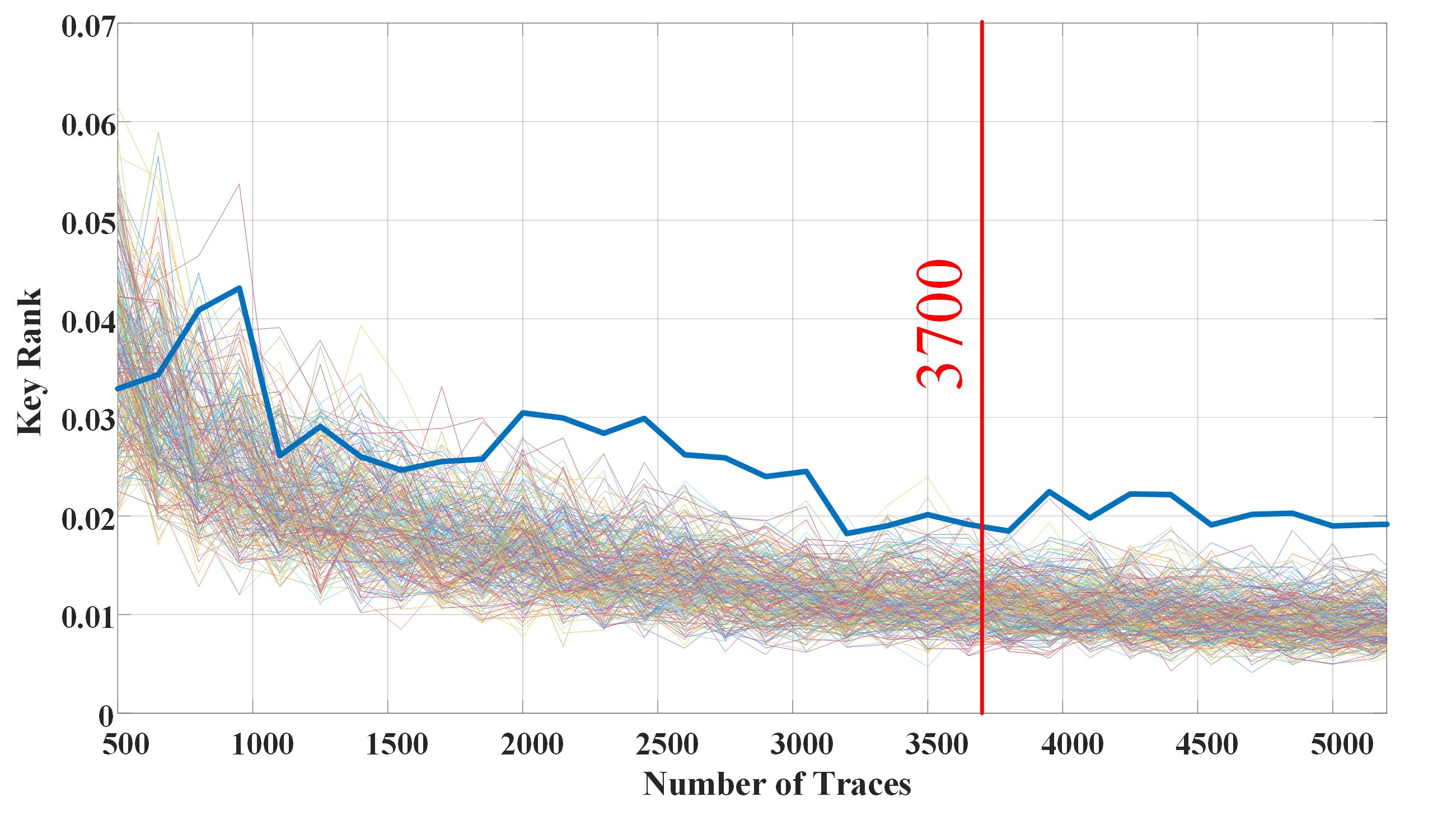}
	\vspace{-0.5cm}
	\caption{Rank of key candidates versus number of traces in a SCAUL attack with a leakage model obtained with sensitivity analysis.}
	\label{fig:scaul_result}
\end{figure}

The data-dependent features of power measurements extracted by the auto-encoder can also be used to identify the leakage model efficiently. We repeat a similar DPA attack with results shown in Fig. \ref{fig:dpa} (b) but this time we employ the leakage model identified by the sensitivity analysis of Section \ref{sec:sensitivity} instead of the HW model. The rank of key candidates versus the number of power measurements is shown in Fig. \ref{fig:scaul_result}. We notice that the correct key takes the highest rank if at least 3700 measurements (encryptions) are available. This is a degradation of around $2.3\times$ in efficiency compared to a model-based attack.

We point out that the power features used in the SCAUL attack with the results in Fig. \ref{fig:scaul_result} are the same features as in Fig. \ref{fig:dpa} (b). The larger amount of measurements required in SCAUL compared to a model-based attack is the cost of detecting a proper leakage model. In other words, if a prior information is available, it can be used to achieve higher efficiency with the auto-encoder features. Otherwise, more measurements are required to retrieve the information.

\subsection{Misaligned Traces}
In addition to unsupervised leakage detection and significant efficiency improvement, another major advantage of SCAUL over classical power analysis techniques is the ability to recover the secret key even with non-synchronous measurements. 

In this experiment, we let the measurement windows for extracting power traces of individual S-box operations, as shown in Fig. \ref{fig:horizon}, take on a random shift. Specifically, we locate the windows on positions $r_i+s_i,i=0,1,\cdots, 15$ in which $s_i$'s are uniformly distributed random variables in $[-12.5, 12.5]$ and $r_i$'s are the accurate positions of S-box operations. It simulates a scenario in which the timing of the measurements is imprecise with an uncertainty equal to $20\%$ of the hardware clock cycle. We set the length of the windows to $150$ samples to cover the entire duration of a S-box operation amid the misalignment. 

\begin{figure}[t!]
	\centering
	\includegraphics[width=0.4\textwidth]{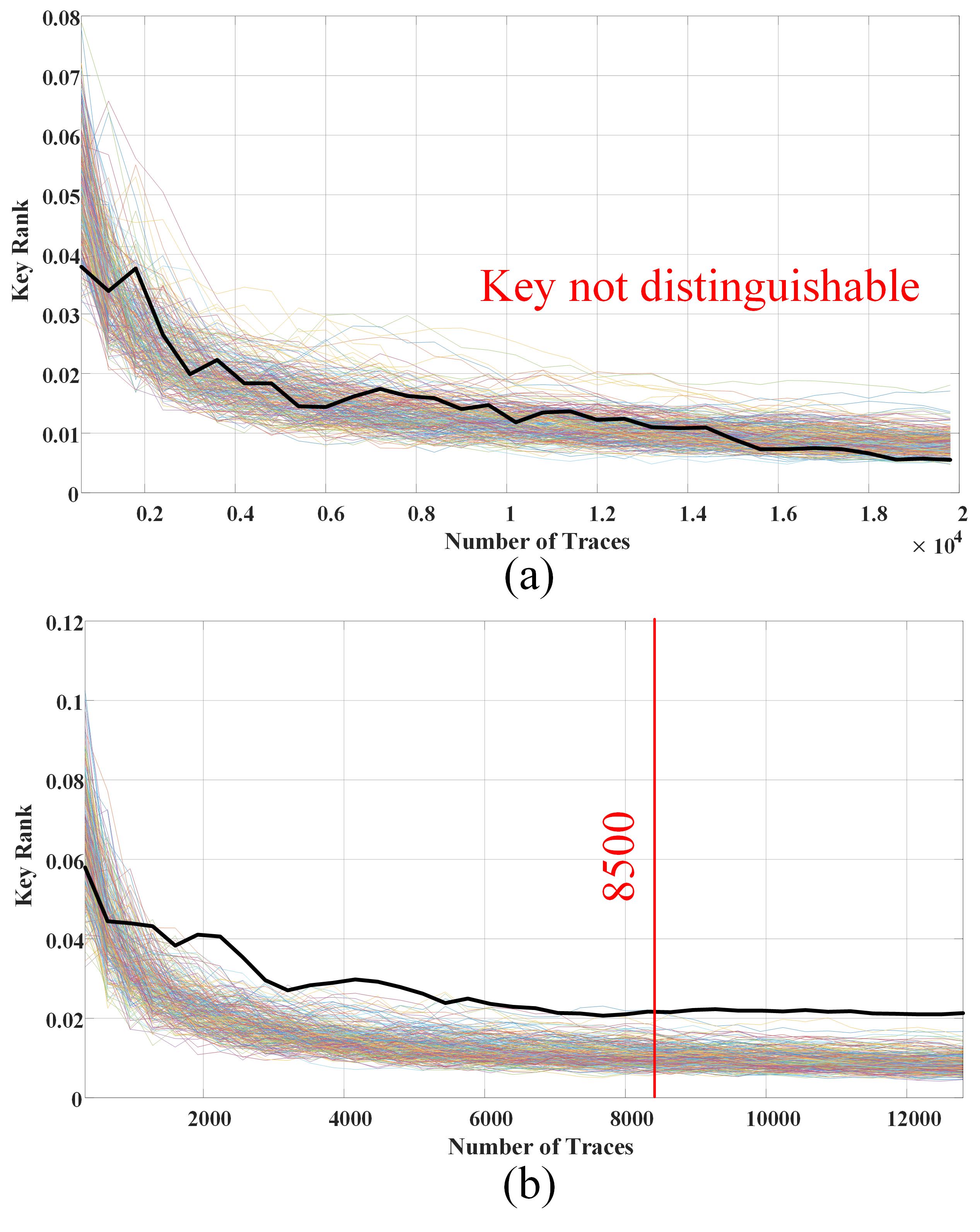}
	\vspace{-0.4cm}
	\caption{Rank of key candidates versus number of traces in a DPA attack with HW leakage model and misaligned traces; (a) maximum difference of power samples, (b) maximum difference of power features.}
	\label{fig:misalign}
\end{figure}

The result of a classical DPA attack with HW model and misaligned traces is shown in Fig. \ref{fig:misalign} (a). As expected the correct key is not distinguishable with 20K measurements since classical techniques assume data-dependent features appear at the same time sample. However, by using the auto-encoder power features, the correct key is recovered with only 8500 measurements as shown in Fig. \ref{fig:misalign} (b). This experiment demonstrates that the data-dependent features are encoded into the internal representation of the auto-encoder even if they are spread over different time samples.

\begin{figure}[t!]
	\centering
	\includegraphics[width=0.4\textwidth]{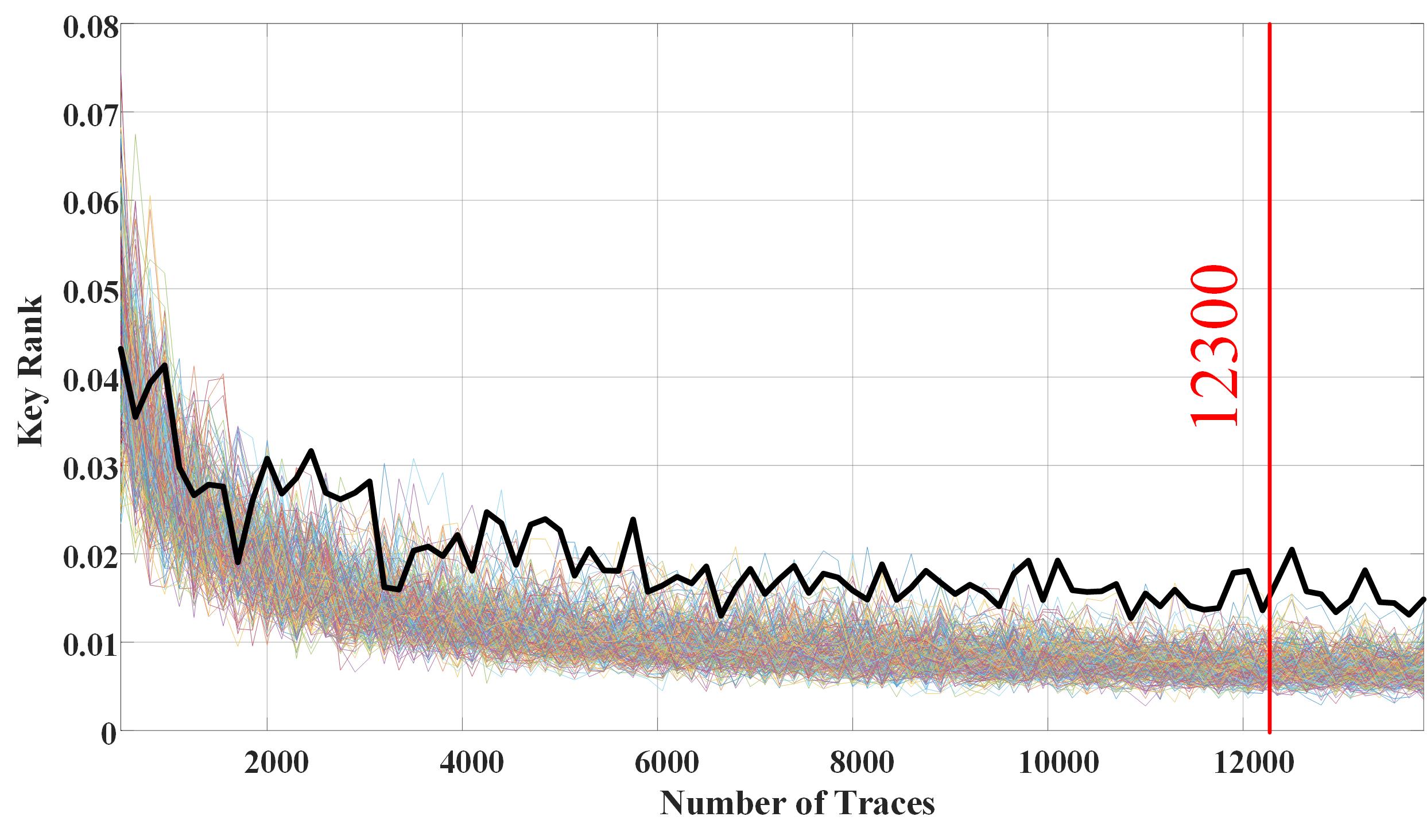}
	\vspace{-0.5cm}
	\caption{Rank of key candidates versus number of traces in a SCAUL attack with a leakage model obtained with sensitivity analysis and misaligned power traces.}
	\label{fig:scaul_misalign}
\end{figure}

The result of an SCAUL attack with misaligned measurements and sensitivity analysis for leakage model detection is shown in Fig. \ref{fig:scaul_misalign}. It is observed that even without using prior knowledge of a leakage model and with misaligned traces, SCAUL is able to recover the correct key with 12300 measurements. Hence, the extracted features of the auto-encoder contain all information about data-dependent samples of the power traces sufficient for leakage detection.

\section{Conclusions} \label{sec:conclusion}
We introduced an unsupervised learning approach for side-channel analysis, called SCAUL, capable of extracting information about data processed on hardware without requiring prior knowledge on the leakage model or training data. 
At the heart of SCAUL, there is an auto-encoder that encodes the data-dependent samples of the power measurements into a neural representation with the highest mutual information with the secret data, and which is used for identifying a proper leakage model using sensitivity analysis.
On a lightweight implementation of AES on Artix-7 FPGA, we demonstrated that an LSTM auto-encoder can improve the efficiency of a classical model-based DPA attack by $10\times$. We also showed that SCAUL is able to identify a proper leakage model from the auto-encoder features and recover the correct key with less than 3700 measurements, compared to 17400 traces required in a DPA attack. With imprecise measurements in which the timing uncertainty is around 20\% of the hardware clock cycles, SCAUL can still recover the secret key with 12300 measurements while classical DPA fails to detect the key with more than 20K traces.

\ifCLASSOPTIONcaptionsoff
  \newpage
\fi


\bibliographystyle{IEEEtran}
\bibliography{IEEEabrv,bibliography.bib}
\end{document}